\documentclass[aps,prd,unsortedaddress,superscriptaddress,showpacs,nofootinbib,11pt]{revtex4-1}
\usepackage{graphicx,color}
\usepackage{relsize}
\usepackage{slashed}
\usepackage[normalem]{ulem}
\usepackage{tabu}
\usepackage{rotating}
\usepackage{bigstrut}
\usepackage{makecell}
\usepackage[dvipsnames]{xcolor}
\usepackage{rotate}
\usepackage{longtable}
\usepackage{amsmath}
\usepackage{amssymb,bm}
\usepackage{amsthm}
\usepackage{mathrsfs}
\usepackage{array}
\usepackage[all]{xy}
\usepackage{euscript}
\usepackage{enumerate}
\usepackage{slashed}
\usepackage{appendix}
\usepackage{float}

\usepackage{subfigure} % NEEDED IN ORDER TO USE SUBFIG
\usepackage{mathtools}

\usepackage{soul}
\usepackage{hyperref}
\hypersetup{pdftex,colorlinks=true,linkcolor=blue,citecolor=blue,menucolor=black,urlcolor=blue,filecolor=blue}

\newcommand{\wuhao}[1]{{\color{purple} #1}}
\begin{document}
	\title{The low-lying light tetraquark states with quantum numbers $J^{P}=0^{+ }$, $1^{+}$ and $2^{+}$ }
	
	\author{Hao Wu}
	\affiliation{School of Physical Science and Technology, Southwest
		University, Chongqing 400715, China}

  \author{Mao-Jun Yan}\email{yanmj0789@swu.edu.cn}
	\affiliation{School of Physical Science and Technology, Southwest
		University, Chongqing 400715, China}

	\author{Chun-Sheng An}\email{ancs@swu.edu.cn}
	\affiliation{School of Physical Science and Technology, Southwest
		University, Chongqing 400715, China}

 \author{Cheng-Rong Deng}\email{rdeng@swu.edu.cn}
	\affiliation{School of Physical Science and Technology, Southwest
		University, Chongqing 400715, China}
	
	\date{\today}
	
	\begin{abstract}
			The low-lying light tetraquark states are investigated in the non-relativistic quark model (NRQM) including the pseudoscalar meson exchange,
            where two different confinement potential schemes, the Cornell potential and the linear potential, are employed, along with the instanton-induced interaction serving as the residual spin-dependent interaction. The numerical results show agreement with masses of $f_{0}(500)$, $f_{0}(1370)$, $f_{0}(1500)$, $f_{0}(2020)$, $f_{0}(2200)$, $h_{1}(1170)$, $h_{1}(1595)$, $h_{1}(1900)$, $h_{1}(1965)$, $h_{1}(2215)$, $f_{2}(1430)$, $f_{2}(1640)$, $f_{2}(1810)$, $f_{2}(2010)$, $f_{2}(2150)$, $a_{0}(980)$, $a_{0}(1450)$, $a_{0}(1950)$, $a_{1}(1260)$, $a_{1}(1640)$, $a_{2}(1700)$, $K^{*}_{0}(1430)$, $K^{*}_{0}(1950)$, $K_{1}(1270)$, $K_{1}(1440)$, $K_{1}(1650)$, and $K^{*}_{2}(1980)$. The results shed light on the spectrum of these mesons and offer guidande to search for the tetraquarks in the future.
	\end{abstract}
	
\maketitle
\section{Introduction.}\label{Introduction}

Since $X(3872)$~\cite{Belle:2003nnu} and $D_{s0}^{\ast}(2317)$~\cite{Belle:2003guh} have been reported in 2003, the physical observation of many new hadronic states is challenging our current understanding of hadrons as conventional mesons and baryons with valence contents of quark-antiquark and three quarks, respectively, since most of them do not fit in the well-known quark model. This difficulty brought back a long-standing discussion on the exotic hadronic structures, i.e., multiquark configurations that might have quantum numbers beyond those assigned to the conventional mesons and baryons.

 $X(3872)$ and $D_{s0}^{\ast}(2317)$  have intrigued the particle physics community due to their unique properties and the insights into the nature of strong interactions.
Therefore, tetraquark candidates in the light sector, cousins of $X(3872)$ and $D_{s0}(2317)$, play a crucial role in understanding the strong interaction in the nonperturbative region. There are dozens of light mesons with masses below $2000\,\rm{MeV}$, where the rich spectrum covers the genuine mesons, glueballs, hybrids, molecules, compact tetraquarks, and admixtures. The candidates of tetraquarks are essential to explore the dynamics of multi-quark states.

The concept of tetraquarks ($qq\bar{q}\bar{q}$) arose from discrepancies in the meson spectrum, such as the overpopulation of states and anomalous properties (e.g. mass, decay width). Light tetraquarks propose bound states of two quarks and two antiquarks, extending beyond the $q\bar{q}$ paradigm. These states may exhibit exotic quantum numbers (e.g. $J^{PC}=0^{--}$, $J^{PC}=1^{-+}$) forbidden by conventional $q\bar{q}$ picture, or manifest as resonant structures in decay channels (e.g. $f_0(500)$, $a_0(980)$).

Jaffe firstly pioneer the systematic study of light tetraquarks, proposing the $qq\bar{q}\bar{q}$ structure for scalar mesons~\cite{Jaffe:1976ig}. Subsequently, Weinstein and Isgur proposed the meson-meson molecular interpretations~\cite{Weinstein:1983gd}, then Maiani $et.~al.$ further extended the diquark-antidiquark model to the light and heavy sectors~\cite{Maiani:2004vq}. In addition, 't Hooft $et.~al.$ have also discussed scalar mesons as tetraquarks in the context of chiral symmetry~\cite{tHooft:2008rus}. Thereafter, in order to get better understanding of light tetraquark states, numerous studies have been conducted to calculate their spectrum using different theoretical approaches, such the QCD sum rule~\cite{Chen:2007xr,Wang:2019nln,Wang:2024pgy}, the relativistic diquark-antidiquark picture~\cite{Ebert:2008id}, the flux-tube model with a multi-body confinement interaction~\cite{Deng_2012}, the modified Godfrey-Isgur (MGI) quark model and quark-pair creation (QPC) model~\cite{Wang:2024lba}, and the lattice QCD~\cite{Zhao:2021jss}.

In the scalar sector, there are low-lying $f_0(500)$, $\kappa(700)$, $f_0(980)$, and $a_0(980)$, which are widely studied in models. The details of those states can be referred to in Refs.~\cite{Zou:1993az,Zou:1994ea,Locher:1997gr,Zheng:2003rw,Pelaez:2020gnd,Cao:2024zuy} and the references therein.
In the region of $1700-2000 \,\rm{MeV}$,
$a_0(1710)$, with mass and width of \(1704\pm5\pm2\,\mathrm{MeV}\) and \(110\pm15\pm11 \,\rm{MeV}\), is reported in $\eta\pi$ invariant mass distributions in $\eta_c\to \eta\pi\pi$~\cite{BaBar:2021fkz}. $a_0(1710)$, $f_0(1710)$, and $f_0(1770)$ are confirmed in $K\bar{K}$ invariant mass distributions in $D_s\to K_s K \pi$~\cite{BESIII:2022npc}.
Additionally,
$f_0(2020)$ and $f_0(2470)$ are reported in the $J/\psi \to \gamma \eta^{\prime}\eta^{\prime}$~\cite{BESIII:2022zel}.

The light axial vector particles listed on the PDG~\cite{ParticleDataGroup:2024cfk} are in the region of $1100-1600$~MeV, where $f_1(1285)$, $f_1(1420)$, and $f_1(1510)$ are reported in $K^{0}_{s}K^{0}_{s}\pi^{0}$ invariant mass in $J/\psi\to\gamma K^{0}_{s}K^{0}_{s}\pi^{0}$~\cite{BESIII:2022chl}. $h_1(1170)$ is observed in the $3\pi$ invariant mass in $\pi p\to3\pi n$~\cite{Dankowych:1981ks,Ando:1990ti}, $h_1(1415)$ is observed in the $K\bar{K}\pi$ invariant mass in $K^-p\to K^0_sK^\pm\pi^\mp\Lambda$~\cite{Bityukov:1986yd} as well as in the $\eta^{\prime}\gamma$ invariant mass distributions in $J/\psi\to\gamma\eta'\eta'$~\cite{BESIII:2022zel}, and $h_1(1595)$ is reported in the $\omega\eta$ invariant mass distributions in $\pi^- p\to\omega\eta n$~\cite{BNL-E852:2000poa}. $a_1(1260)$ and $a_1(1640)$ are observed in diffractive dissociation of $190\,\rm{GeV}$ pions into $\pi^-\pi^+\pi^-$~\cite{JPAC:2018zwp}. $b_1(1235)$ is reported in the $\omega\pi$ channel in $p\pi$ reaction~\cite{Abolins:1963zz}. Additionally, BESIII reports a new $h_1(1900)$ meson by analyzing the $\eta\pi$ invariant mass distributions in $J/\psi \to \omega\eta$~\cite{besiiicollaboration2023studydecayjpsito}. COMPASS collaboration identifies a new $a_1(1420)$ meson with mass and width of \(1414^{+15}_{-13}\,\mathrm{MeV}\) and \(153^{+8}_{-23} \,\rm{MeV}\) in diffractive dissociation of $190\,\rm{GeV}$ pions into $\pi^-\pi^+\pi^-$~\cite{COMPASS:2020yhb}. Meanwhile, Crystal Barrel collaboration first observes $h_1(1965)$ and $h_1(2215)$ in $\omega\eta$ invariant mass distributions in $p\bar{p}\to \omega\eta$~\cite{Anisovich_2002}.

In the tensor sector, there are twelve light isoscalar tensor mesons with $I=0$ listed in PDG~\cite{ParticleDataGroup:2024cfk}, $i.e.$, $f_2(1270)$, $f_2(1430)$, $f_2(15525)$, $f_2(1565)$, $f_2(1640)$, $f_2(1810)$, $f_2(1910)$, $f_2(1950)$, $f_2(2010)$, $f_2(2150)$, $f_2(2300)$, $f_2(2340)$. These resonances are intensively studied in radiative $J/\psi$ decays~\cite{BES:1999dmf,Ablikim_2006,Ablikim_2013,Ablikim_2013(2),Ablikim_2015,Aaltonen_2016,Ablikim_2018,Ablikim_2020,Bugg:2009ch,Chen_2022}. Tensor mesons with $I=1$, $a_2(1320)$ and $a_2(1700)$, are reported in diffractive dissociation of $\pi p$ collisions \cite{Conte:1967zzb}, in two-photon collisions \cite{L3:1997mpi}, in $p\bar{p}$ collisions \cite{CrystalBarrel:2002qpf,Anisovich:2009zza,Kopf:2020yoa}.

To understand the mesons above, the spectrum is widely studied in phenomenological methods. In addition to the studies on the compact tetraquarks in the classical quark potential model~\cite{Yang:2009zzp,Deng:2016rus,Richard:2017una,Luo:2017eub,Jin:2020yjn,Yang:2021zhe,Deng:2021gnb}, the QCD sum rule approach~\cite{Chen:2013pya,Chen:2015ata,Wang:2017dtg,Wang:2019got}, and the diquark-diquark framework~\cite{Ebert:2007rn,Esposito:2013fma,Shi:2021jyr}, the interactions driven by instanton are introduced to the S-wave tetraquark spectrum in the non-relativistic quark model (NRQM). The Cornell potential~\cite{Eichten:1974af,Wang:2022clw,Zhang:2022qtp} and linear potential~\cite{Beinker:1995qe} are considered, where the instanton-induced interaction serves as the residual spin-dependent interaction ~\cite{Shuryak:1988bf,Blask:1990ez,Brau:1998sxe,Semay:2001th,Beinker:1995qe,tHooft:2008rus,An:2013zoa,Wang:2022clw,Zhang:2022qtp,Jiang:2024hxq}.

This article is organized as follows. Sect.~\ref{Framework} presents the formalism of our model, including effective Hamiltonian, wave function configurations, and model parameters. Numerical results and discussions for the spectrum of fully-light tetraquark states in two models are shown in Sect.~\ref{Results and Discussions}. Finally, Sect.~\ref{General Discussion and Conclusion} contains a brief conclusion.

\clearpage

\section{Theoretical Framework}\label{Framework}
\subsection{Effective Hamiltonian}\label{Effective Hamiltonian}
In this study, the non-relativistic quark potential model is utilized to explore the spectrum of the tetraquark states. In the said model, the effective Hamiltonian reads
	\begin{equation}
		H_{eff.}=\sum_{i=1}^{4}\left(m_i+T_i\right)-T_{C.M.}+V_{Conf.}+V_{Ins.}+V_{OBE.},
  \label{heff}
	\end{equation}
where $m_i$ and $T_i$ are the constituent mass and kinetic energy of $i$-th quark, respectively. $T_{C.M.}$ is the energy in the center of mass frame (C.M.). $V_{Conf.}$ stands for the quark confinement potential, corresponding to the Cornell potential~\cite{Eichten:1974af,Wang:2022clw,Zhang:2022qtp} and the linear potential~\cite{Beinker:1995qe}, respectively,
\begin{equation}
V_{Cornell.}=\sum_{i<j}-\frac{3}{16}\,\left(\vec{\lambda}^c_i\cdot\vec{\lambda}^c_j\right)\,\left(b\,r_{ij}-\frac{4}{3}\frac{\alpha_{ij}}{r_{ij}}+C_0\right)\,
\label{cornell}
\end{equation}
and
\begin{equation}
V_{Linear.}=\sum_{i<j}-\frac{3}{16}\,\left(\vec{\lambda}^c_i\cdot\vec{\lambda}^c_j\right)\,\left(a_{ij}+b\,r_{ij}\right),
	\end{equation}
where $\vec{\lambda}^{c}_{i(j)}$ is Gell-Mann matrix in $SU(3)$ color space acting on the $i(j)$-th quark, $b$, $\alpha_{ij}$ and $C_0$ are strength of quark confinement, QCD running coupling constant between two quarks and zero point energy, respectively. $a_{ij}, b_{ij}$ are the offset and slope of relative quark–(anti)quark distance. Notice that the choices of $a_{ij}\,\left( b_{ij}\right)$ are different with respect to the diquarks of $qq$ and $q\bar{q}$.
	
For the residual spin-dependent interaction, the 't Hooft's instanton-induced interaction~\cite{Migura:2006ep} is employed,
	\begin{eqnarray}
V_{Ins.}=V^{qq}_{Ins.}+V^{q\bar{q}}_{Ins.},\\
	\end{eqnarray}
	with
	\begin{align}
		V^{qq}_{Ins.}=&\sum_{i<j}-\hat{g}^{qq}_{ij}\left(P_{ij}^{S=1}P_{ij}^{C,\bf{6}}+2P_{ij}^{S=0}P_{ij}^{C,\bf{\bar{3}}}\right)\delta^3\left(\vec{r}_{ij}\right),\\
		V^{q\bar{q}}_{Ins.}=&\sum_{i<j}\hat{g}^{q\bar{q}}_{ij}\left[\frac{3}{2}P_{ij}^{S=1}P_{ij}^{C,\bf{8}}+P_{ij}^{S=0}\left(\frac{1}{2}P_{ij}^{C,\bf{8}}+8P_{ij}^{C,\bf{1}}\right)\right]\delta^3\left(\vec{r}_{ij}\right),
	\end{align}
where $V^{qq}_{Ins.}$ acts both on quark-quark and antiquark-antiquark pair in addition to $V^{q\bar{q}}_{Ins.}$ acting on quark-antiquark pair. $\hat{g}^{qq}_{ij}$ and $\hat{g}^{q\bar{q}}_{ij}$ are flavor-dependent coupling strength operators~\cite{Wang:2022clw}. The spin projection operators $P_{ij}^{S=0}$, $P_{ij}^{S=1}$ correspond to spin-singlet and spin-triplet states, respectively. So do $P_{ij}^{C,\bf{\bar{3}}}$, $P_{ij}^{C,\bf{6}}$, $P_{ij}^{C,\bf{1}}$ and $P_{ij}^{C,\bf{8}}$ acting on the color space with the superscripts labeling color antitriplet $\bf{\bar{3}}_c$, color sextet $\bf{6}_c$, color singlet $\bf{1}_c$ and color octet $\bf{8}_c$, respectively.

Concerning the contact interaction in the form of the Dirac delta function $\delta^{3}(\vec{r}_{ij})$ causes problems in numerically solving the eigenequation of the Hamiltonian, the delta function is regularized by a Gaussian regulator~\cite{Godfrey:1985xj,Vijande:2004he,Beinker:1995qe},
	\begin{eqnarray}
		\delta^3\left(\vec{r}_{ij}\right)\,\rightarrow\,\left(\frac{\sigma}{\sqrt{\pi}}\right)^3\mathrm{exp}\left(-\sigma^2\,r_{ij}^2\right).
	\end{eqnarray}
where $\sigma$ is a regularization parameter originating from the finite sizes of the constituent quarks.

In addition to the gluon exchange between quarks in short range, the Goldstone boson (pseudoscalar meson) exchange potentials provide the long-range quark-quark interactions, with the following explicit functional forms~\cite{fujiwara1996effective,fujiwara2001interactions,Vijande:2004he},
\begin{eqnarray}
V_{OBE.}=\sum_{i<j}\left(V_{ij}^{\pi}+V_{ij}^{K}+V_{ij}^{\eta}\right),
	\end{eqnarray}
    with
	\begin{align}
		V_{ij}^{\pi}=&\frac{g^{2}_{ch}}{4\pi}\frac{m^{2}_{\pi}}{12m_{i}m_{j}}\frac{\Lambda^{2}_{\pi}m_{\pi}}{\Lambda^{2}_{\pi}-m^{2}_{\pi}}\,\vec{\sigma}_i\cdot\vec{\sigma}_j\left[Y(m_{\pi}\vec{r}_{ij})-\frac{\Lambda^{3}_{\pi}}{m^3_{\pi}}Y(\Lambda_{\pi}\vec{r}_{ij})\right]\sum_{a=1}^{3}\left(\vec{\lambda}^a_i\cdot\vec{\lambda}^a_j\right),\\
        V_{ij}^{K}=&\frac{g^{2}_{ch}}{4\pi}\frac{m^{2}_{K}}{12m_{i}m_{j}}\frac{\Lambda^{2}_{K}m_{\pi}}{\Lambda^{2}_{K}-m^{2}_{K}}\,\vec{\sigma}_i\cdot\vec{\sigma}_j\left[Y(m_{K}\vec{r}_{ij})-\frac{\Lambda^{3}_{K}}{m^3_{K}}Y(\Lambda_{K}\vec{r}_{ij})\right]\sum_{a=4}^{7}\left(\vec{\lambda}^a_i\cdot\vec{\lambda}^a_j\right),\\
        V_{ij}^{\eta}=&\frac{g^{2}_{ch}}{4\pi}\frac{m^{2}_{\eta}}{12m_{i}m_{j}}\frac{\Lambda^{2}_{\eta}m_{\eta}}{\Lambda^{2}_{\eta}-m^{2}_{\eta}}\,\vec{\sigma}_i\cdot\vec{\sigma}_j\left[Y(m_{\eta}\vec{r}_{ij})-\frac{\Lambda^{3}_{\eta}}{m^3_{\eta}}Y(\Lambda_{\pi}\vec{r}_{ij})\right]\left[cos\theta_{p}\left(\vec{\lambda}^8_i\cdot\vec{\lambda}^8_j\right)-sin\theta_{p}\right],
        \label{mexch}
	\end{align}
where the $\vec{\lambda}_{i(j)}$ is the SU(3) flavor Gell-Mann matrices. The $\Lambda_{\pi(K,\,\eta)}$ and $\theta_{p}$ are the cutoff parameters and mixing angle, which are quoted from Ref.~\cite{Vijande:2004he}. The $m_{\pi}$, $m_{K}$, and $m_{\eta}$ are the physical masses of the SU$\left(3\right)$ Goldstone bosons. $Y(x)$ is defined by $Y(x)=e^{-x}/x$. The chiral coupling constant $g_{ch}$ can be obtained from the $\pi NN$ coupling constant through Ref.~\cite{Scadron:1982eg}
\begin{equation}
		\frac{g^{2}_{ch}}{4\pi}=\left(\frac{3}{5}\right)^{2}\frac{g^{2}_{\pi NN}}{4\pi}\frac{m^{2}_{u,d}}{m^{2}_{N}}.
	\end{equation}

	\subsection{Wave Functions}\label{Wave Functions}
	The wave function of a tetraquark state is a product of the orbital, color, spin, and flavor wave functions,
	\begin{equation}
	\Psi=\psi_{orbital}\otimes\psi_{color}\otimes\psi_{spin}\otimes\psi_{flavor}.
	\end{equation}
	The orbital wave function of a few-body system is expanded in terms of a set of Gaussian basis functions, forming an approximated complete set in a finite coordinate space~\cite{Hiyama:2003cu}. The wave function of the ground four-quark states in the coordinate space reads,
	\begin{equation}
		\psi_{orbital}(\{\vec{r}_i\})=\prod_{i=1}^{4}\sum_{j}^{n}C_{ij}\left(\frac{1}{\pi \beta_{ij}^2}\right)^{3/4}\mathrm{exp}\left[-\frac{1}{2\beta_{ij}^2}r_i^2\right],
	\end{equation}
	where $\{\beta_{ij}\}$ are the harmonic oscillator length parameters, and originated from the angular frequencies $\{\omega_{j}\}$ of the harmonic oscillator by $1/\beta_{ij}^2=m_i\omega_{j}$. Meanwhile, we assume that $\{\omega_{j}\}$ are independent of the quark masses, $m_i$, i.e., then the spatial wave function can be simplified as,
	\begin{align}
		\psi_{orbital}(\{ \vec{r}_i \})=&\sum_{j}^{n}C_{j}\prod_{i=1}^{4}\left(\frac{m_i\omega_{j}}{\pi }\right)^{3/4}\mathrm{exp}\left[-\frac{m_i\omega_{j}}{2}r_i^2\right]\notag\\	=&\sum_{j}^{n}C_{j}\,\varphi\left(\omega_{j},\{\vec{r}_i\}\right),\label{SpatialF}
	\end{align}
	which is always adopted for the calculations of multiquark systems~\cite{Zhang:2007mu,Zhang:2005jz,Liu:2019zuc}.

    Following the work in Ref.~\cite{Hiyama:2003cu}, we define $1/\beta_{ij}^2=1/\beta_{j}^2=m_q\omega_{j}$ ($m_q$ denotes the constituent mass of the corresponding quark), thus the parameters $\beta_{j}$ are set to be geometric series,
	\begin{equation}
		\beta_{j}=\beta_1a^{j-1}\hspace{0.8cm}\left(j=1,2,...,n\right),
	\end{equation}
	where the parameters $\{\beta_{1},\beta_{n},n\}$ are the Gaussian size parameters in geometric progression for numerical calculations, and the final results are stable and independent of these parameters within an approximated complete set in a large space.
	
	In addition, to remove the influence of the center-of-mass kinetic energy, here we rewrite the tetraquark wave function using the coordinates defined as follows,
    \begin{align}
		\vec{\xi}_{1}=&\vec{r}_{1}-\vec{r}_{2},\\
		\vec{\xi}_{2}=&\vec{r}_{3}-\vec{r}_{4},\\
        \vec{\xi}_{3}=&\frac{m_1\vec{r}_{1}+m_2\vec{r}_{2}}{m_1+m_2}-\frac{m_3\vec{r}_{3}+m_4\vec{r}_{4}}{m_3+m_4},\\
        \vec{\xi}_{4}=&\frac{m_1\vec{r}_{1}+m_2\vec{r}_{2}+m_3\vec{r}_{3}+m_4\vec{r}_{4}}{m_1+m_2+m_3+m_4},\label{Jacobi coordinates}
	\end{align}
with these one can rewritten the Eq.~\eqref{SpatialF}, as
	\begin{equation}
		\psi_{orbital}(\{\vec{\xi}_i\})=\sum_{j}^{n}C_{j}\prod_{i=1}^{4}\left(\frac{\mu_i\omega_{j}}{\pi }\right)^{3/4}\mathrm{exp}\left[-\frac{\mu_i\omega_{j}}{2}\vec{\xi}_i^2\right],
  \label{Spatial Eq}
	\end{equation}
where $\mu_1\equiv\frac{m_1m_2}{m_1+m_2}$, $\mu_2\equiv\frac{m_3m_4}{m_3+m_4}$, $\mu_3\equiv\frac{(m_1+m_2)(m_3+m_4)}{M}$, $\mu_4=M\equiv m_1+m_2+m_3+m_4$. With
the trail wave function defined in Eq.~\eqref{Spatial Eq}, the kinetic energy matrix element is worked out to be
\begin{equation}
		\langle \sum_{i=1}^{4}T_i-T_{C.M} \rangle=\frac{9}{4}\sum_j^n\sum_k^nC_jC_k\frac{(\omega_j\omega_k)^4}{(\frac{\omega_j+\omega_k}{2})^7}.
  \label{elmd}
	\end{equation}
Note that the coordinate $\vec{\xi}_{4}$ should indicate motion of the center of mass of the studied tetraquark state, and it's of course that $\vec{\xi}_{4}=0$ in the C.M. frame. While in present case, the above equations including the used Hamiltonian and wave functions are not written as the forms in the tetraquark C.M. frame, and contributions from motions of the center of mass are finally eliminated. As shown in Eq.~(\ref{heff}), the kinetic energy of the C.M. can be subtracted by the term $-T_{C.M}$, which leads to the final kinetic energy given in Eq.~(\ref{elmd}). On the other hand, one may doubt that the nonzero coordinate $\vec{\xi}_{4}$ should result in additional non-vanishing matrix elements of the inner quark potential, but note that all the potential terms in Eqs.~(\ref{cornell}-\ref{mexch}) are functions of the relative coordinates of the interacting quark pairs, namely $\vec{r}_{ij}=\vec{r}_{i}-\vec{r}_{j}$, and by re-solving Eqs.~\ref{Jacobi coordinates}, one can get that $\{\vec{r}_{i}\}$ can be expressed by $\{\vec{\xi}_{i}\}$ as follow:
\begin{equation}
\vec{r}_{i}=-\vec{\xi}_{4}+\sum_{m=1}^{3}a_{im}\vec{\xi}_{m}\,,
\end{equation}
with $a_{im}$ being corresponding constants which can be expressed by the quark masses. Thus, all the potential terms for any interacting quark pairs in the tetraquark system are independent on the coordinate $\vec{\xi}_{4}$ although $\vec{\xi}_{4}\neq0$, since $\vec{r}_{ij}$ doesn't depend on $\vec{\xi}_{4}$. Consequently, contributions from motions of C.M. of the studied tetraquark systems could be eliminated completely by presently used scheme.

	Stemming from the Pauli principle, color confinement for the tetraquark system, and the ground state is characterized by a symmetric spatial wave function, we construct eight symmetric configurations for $qq\bar{q}\bar{q}$ in the space of color$\otimes$spin$\otimes$flavor,
	\begin{flalign}
		\hspace{2.5cm}|1\rangle&=[qq]_{{\bf{6}}_c{\bf{1}}_s{\bf{6}}_f}[\bar{q}\bar{q}]_{\bar{\bf{6}}_c\bf{1}_s\bar{\bf{6}}_f}, & |2\rangle&=[qq]_{\bar{\bf{3}}_c{\bf{1}_s}\bar{\bf{3}}_f}[\bar{q}\bar{q}]_{{\bf{3}}_c{\bf{1}}_s{\bf{3}}_f},\hspace{2.5cm}\notag\\
		|3\rangle&=[qq]_{{\bar{\bf{3}}_c\bf{3}}_s{\bf{6}}_f}[\bar{q}\bar{q}]_{{\bf{3}}_c{\bf{1}}_s{\bf{3}}_f},& |4\rangle&=[qq]_{{\bar{\bf{3}}}_c{\bf{1}}_s\bar{\bf{3}}_f}[\bar{q}\bar{q}]_{{\bf{3}}_c{\bf{3}}_s\bar{\bf{6}}_f}, \nonumber\\
		|5\rangle&=[qq]_{{\bf{6}}_c{\bf{3}_s}\bar{\bf{3}}_f}[\bar{q}\bar{q}]_{\bar{\bf{6}}_c{\bf{1}_s}\bar{\bf{6}}_f},& |6\rangle&=[qq]_{{\bf{6}}_c{\bf{1}}_s\bf{6}_f}[\bar{q}\bar{q}]_{\bar{\bf{6}}_c{\bf{3}_s}{\bf{3}}_f},\nonumber\\
		|7\rangle&=[qq]_{\bar{\bf{3}}_c{\bf{3}_s}{\bf{6}}_f}[\bar{q}\bar{q}]_{{\bf{3}}_c{{\bf{3}}_s}\bar{\bf{6}}_f},& |8\rangle&=[qq]_{{\bf{6}}_c{\bf{3}_s}\bar{\bf{3}}_f}[\bar{q}\bar{q}]_{\bar{\bf{6}}_c{\bf{3}_s}{\bf{3}}_f}, \label{eq:configurations}
	\end{flalign}
where the $q$ stands for $ u, d, s$ quarks. The color structures $\bf{6}$ and $\bar{\bf{3}}$ ($\bar{\bf{6}}$ and $\bf{3}$) of diquarks (antidiquarks) are represented by the $c$-subscript, the spin structures $\bf{3}$ and $\bf{1}$ of the diquarks (antidiquarks) are represented by the $s$-subscript, and the flavor structures $\bf{6}$ and $\bar{\bf{3}}$ ($\bar{\bf{6}}$ and $\bf{3}$) of diquarks (antidiquarks) are represented by the $f$-subscript, respectively. The explicit flavor wave functions of $nn\bar{n}\bar{n}$, $nn\bar{n}\bar{s}$, $ns\bar{n}\bar{s}$, $ns\bar{s}\bar{s}$, $ss\bar{s}\bar{s}$ are presented in the Appendix~\ref{Flavor Wave Functions}.

\subsection{Strategy}
A basic assumption of the non-relativistic quark potential model is that quarks, trapped inside hadrons, are non-relativistic, obeying the Schrödinger Equation~\cite{Manohar:1983md}. According to the matrix elements in Sec.~\ref{Effective Hamiltonian} and ~\ref{Wave Functions}, the mass spectra can be obtained by solving the generalized matrix eigenvalue problem,
			\begin{equation}
				\sum_{j}^n\sum_{k}^nC^i_{j}\left(H^d_{jk}-E_i^d N_{jk}\right)C^i_{k}=0\,,\label{GEP}
			\end{equation}
			where $i=1\text{--}n$ and
			\begin{align}
				H^d_{jk}=&\langle\psi\left(\omega_{j}\right)(CSF)|H_{eff.}|\psi\left(\omega_{k}\right)(CSF)\rangle,\\
				N_{jk}=&\langle\psi\left(\omega_{j}\right)(CSF)|\psi\left(\omega_{k}\right)(CSF)\rangle.
			\end{align}
The $H^d_{jk}$ are the matrix elements in the total spin-color-flavor-spatial bases, $E_i^d$ stands for the eigenvalues, $C^i_{k}$ are the corresponding eigenvector, and $(CSF)$ stand for color$\otimes$spin$\otimes$flavor wave function. Concerning the Gaussian function associated with different harmonic oscillator frequencies which are not orthogonal, $N_{jk}$ is not an identity matrix. 	

The procedures of the numerical calculations are organized as
firstly, we derive $n$ eigenenergies $E_i^d$ and $n$ corresponding eigenvectors $C^i$ by solving Eq.~\eqref{GEP} to get the masses of single configurations, and then calculate the off-diagonal effects between different configurations. Secondly, according to the Rayleigh-Ritz variational principle, the lowest eigenenergy $E_m^d$ should correspond to the steady state energy. Therefore, we select a set of $\{\beta_{1},\beta_{n},n\}$, then systematically vary and refine the harmonic oscillator length parameter until the energy $E^{d}_{m}$ converges to a stable minimum.

The parameters $\{\beta_{1},\beta_{n},n\}=\{0.02fm,6fm,40\}$ are adopted in the Model I ($V_{Cornell}$) and II ($V_{Linear}$). There are additional parameters with the same choices in the models, where the parameters are
the coupling strengths in INS hyperfine interaction $g_{nb}$ and $g_{sb}$, the regulation parameter of the $\delta$ function $\sigma$, the constituent quark masses $m_n$ and $m_s$.
 Alternatively, there are differences in the behavior of the confinement in models.
 In Model I, the parameters add the QCD effective coupling constants $\alpha_{nn}$, $\alpha_{ns}$, $\alpha_{ns}$, the quark confinement strength $b$, and the zero-point energy $C_0$. For Model II, the parameters add the slope of the relative quark-(anti)quark distance $b_{qq}, b_{q\bar{q}}$, the offset of the relative quark–(anti)quark distance $a_{qq}, a_{q\bar{q}}$.

 \begin{table}
				\caption{Inputs in two models.}\label{Inputs in two models}
				\renewcommand
				\tabcolsep{1.12cm}
				\renewcommand{\arraystretch}{1.0}
				\begin{tabular}{cc|cc}
					\hline\hline
					\multicolumn{2}{c}{\makecell{Model I}}&\multicolumn{2}{c}{\makecell{Model II}}\\\hline
					Parameter(Unit)& Value & Parameter(Unit) & Value\\\hline
					$m_n\,\,(\mathrm{MeV})$   &  $340$   &$m_n\,\,(\mathrm{MeV})$   &  $340$\\
					$m_s\,\,(\mathrm{MeV})$&$511$&$m_s\,\,(\mathrm{MeV})$&$511$\\\hline
					$\sigma\,\,(\mathrm{MeV})$      &  $200$ &$\sigma\,\,(\mathrm{MeV})$      &  $200$ \\\hline
                    $g_{nn}\,\,(\mathrm{10^{-4}MeV^{-2}})$ & $1.090$&$g_{nn}\,\,(\mathrm{10^{-4}MeV^{-2}})$ & $1.090$ \\
                    $g_{ns}\,\,(\mathrm{10^{-4}MeV^{-2}})$ & $0.657$  &$g_{ns}\,\,(\mathrm{10^{-4}MeV^{-2}})$ & $0.662$ \\\hline
					$b\,\,(\mathrm{MeV\cdot fm^{-1}})$   &  $177$  &$b_{qq}\,\,(\mathrm{MeV\cdot fm^{-1}})$   &  $218$ \\
					&&$b_{q\bar{q}}\,\,(\mathrm{MeV\cdot fm^{-1}})$ &$395$ \\\hline
                    $C_0\,\,(\mathrm{MeV})$   &  $-150$  & $a_{qq}\,\,(\mathrm{MeV})$   &  $-393$  \\
					& &   $a_{q\bar{q}}\,\,(\mathrm{MeV})$   &  $-537$  \\\hline
					$\alpha_{nn}$             &  $0.485$ &&\\
					$\alpha_{ns}$             &  $0.488$&&\\
					$\alpha_{ss}$             &  $0.500$ &  &\\
					\hline\hline
				\end{tabular}
			\end{table}
			\begin{table}
				\caption{Masses of the pseudoscalar and the vector mesons on ground states. Rows denoted by “Model I/II” are numerical results estimated in models I/II, and rows denoted by “Exp.” are physical masses in PDG~\cite{ParticleDataGroup:2024cfk}}.\label{Ground state meson spectrum}
				\renewcommand
				\tabcolsep{0.54cm}
				\renewcommand{\arraystretch}{1.0}
				\begin{tabular}{ccccccccc}
					\hline\hline
					$\sout{States} $&$\pi$&$\rho$&$\omega$&$\phi$& $K$&$K^*$&$\eta$&$\eta'$\\\hline
                    $Exp.(\rm{MeV})$&139&770&782&1019&497&892&548&957\\\hline
                     Model I(MeV)&139& 749&739&1019&497&892&570&964\\\hline
                    Model II(MeV)&139& 750&749&1019&496&890&572&972\\
					\hline\hline
				\end{tabular}
			\end{table}

In the current model, the interaction between quarks and antiquarks in tetraquark states is treated as a simple sum of two-body confinement and two-body residual interactions, which is analogous to the description of mesons in constituent-quark models. Consequently, we use the masses of a set of ground-state mesons listed in the PDG~\cite{ParticleDataGroup:2024cfk} to determine the parameters in the models, the explicit numbers of all the parameters are shown in TABLE~\ref{Inputs in two models}, where the spectrum of the low lying pseudoscalar and vector mesons are displayed in TABLE~\ref{Ground state meson spectrum}.
				
\section{Results and Discussions.}\label{Results and Discussions}
According to the configurations in Eq.~\eqref{eq:configurations}, the wave functions of the tetraquarks are represented by the eight configurations, with details organized in Tables~\ref{Configurations of nnnn Systems},~\ref{Configurations of nsns Systems},~\ref{Configurations of nnns Systems}. With the parameters of two models in Table~\ref{Inputs in two models}, the mass spectra of light tetraquark systems are calculated and presented in Tables~\ref{nnnn states},~\ref{nsns states},~\ref{nnns states}. Consider that a physical tetraquark state should be composed of two types of components with $6_{c}\otimes\bar{6}_{c}$ and $\bar{3}_{c}\otimes3_{c}$ color structures, configuration mixing is taken into account during the numerical calculation.  Additionally, the mass spectrua of $nn\bar{n}\bar{n}$, $ns\bar{n}\bar{s}$, $ss\bar{s}\bar{s}$, $nn\bar{n}\bar{}$, and $ns\bar{s}\bar{s}$ are displayed in Fig.~\ref{nnnn picture},~\ref{nsns picture},~\ref{nnns picture}, where the horizontal lines are two-meson thresholds.
			
\subsection{$nn\bar{n}\bar{n}$ System.}
\begin{figure}[htbp]
				\centering
				\includegraphics[width=1.0\textwidth]{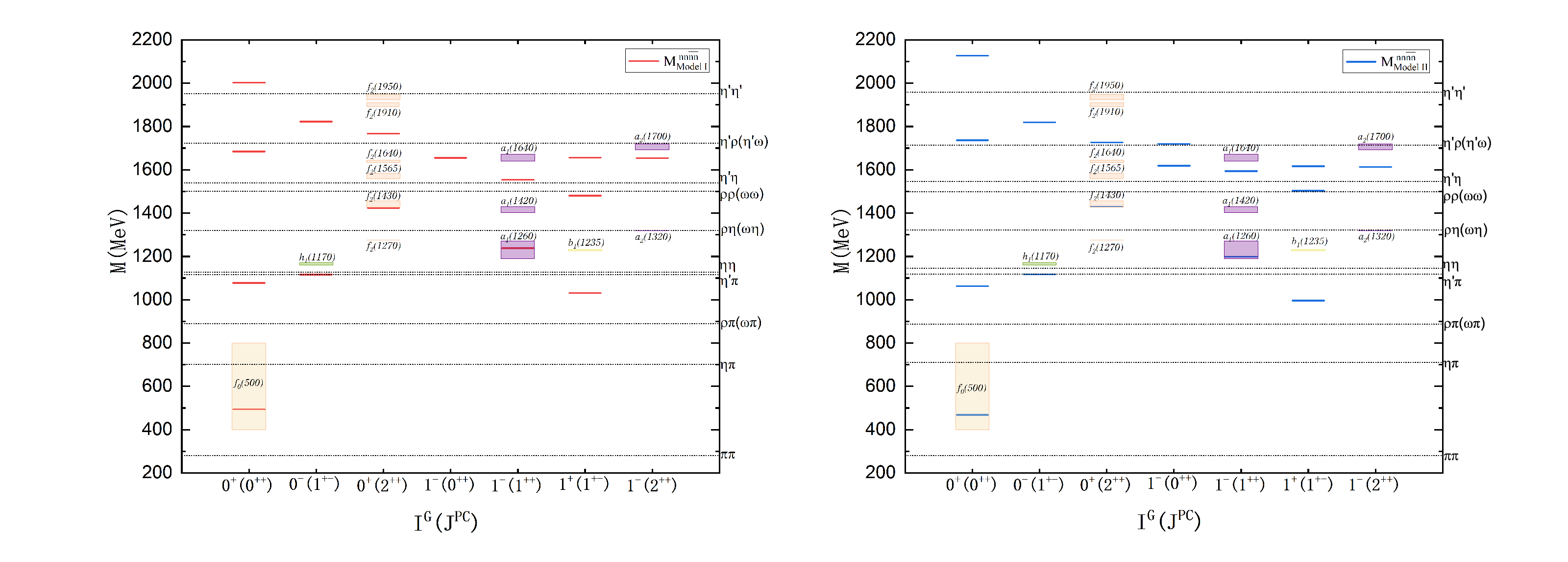}
				\caption{Mass spectrum of S-wave $nn\bar{n}\bar{n}$ states. The data on the left corresponds to Model I, whereas the data on the right pertains to Model II. The red solid lines represent the numerical results of the $nn\bar{n}\bar{n}$ states within Model I, and the blue lines indicate the results for Model II. The gray dotted lines are the two-body thresholds, the rectangles represent the physical masses of $nn\bar{n}s\bar{n}$ quoted from PDG~\cite{ParticleDataGroup:2024cfk}, and the bands of the rectangles stand for the uncertainties of masses.}\label{nnnn picture}
			\end{figure}

The configurations of $nn\bar{n}\bar{n}$ are organized in the Table~\ref{Configurations of nnnn Systems}, where the energy eigenstates are presented in Table~\ref{nnnn states}. Additionally, the mass spectrum of $nn\bar{n}\bar{n}$ is displayed in Fig.~\ref{nnnn picture}.

\begin{table}[htbp]
		\caption{Configurations of $nn\bar{n}\bar{n}$ System.}\label{Configurations of nnnn Systems}
		\renewcommand
		\tabcolsep{0.72cm}
		\renewcommand{\arraystretch}{1.0}
	\begin{tabular}{cccccc}
		\toprule
	$I^{G}$&$J^{PC}$&\multicolumn{4}{c}{Configurations}\\\hline
	$0^{+}$&$0^{++}$&  $|1\rangle$ &$|2\rangle$&$|7\rangle$&$|8\rangle$\\\hline
					$0^{-}$&$1^{+-}$ &$|7\rangle$&$|8\rangle$&&\\\hline
					$0^{+}$&$2^{++}$ &$|7\rangle$&$|8\rangle$&&\\\hline
					$1^{-}$&$0^{++}$&$|1\rangle$&$|7\rangle$& &\\\hline
					$1^{-}$&$1^{++}$&$|3'\rangle=\frac{1}{\sqrt{2}}\left(|3\rangle+|4\rangle\right)$ &  $|5'\rangle=\frac{1}{\sqrt{2}}\left(|5\rangle+|6\rangle\right)$ &&\\\hline
					$1^{+}$&$1^{+-}$& $|4'\rangle=\frac{1}{\sqrt{2}}\left(|3\rangle-|4\rangle\right)$ &   $|6'\rangle=\frac{1}{\sqrt{2}}\left(|5\rangle-|6\rangle\right)$ &  $|7\rangle$  & \\\hline
					$1^{-}$&$2^{++}$ &$|7\rangle$ &&&\\\hline		
					$2^+$ &$0^{++}$& $|1\rangle$&$|7\rangle$&&\\\hline
					$2^{-}$&$1^{+-}$&  $|7\rangle$ &&&\\\hline
					$2^{+}$&$2^{++}$&$|7\rangle$&&&\\		
					\toprule
				\end{tabular}
			\end{table}	

            \begin{table*}[htbp]
				\caption{Numerical results for the $nn\bar{n}\bar{n}$ states in two models. The quantum numbers $I^{G}(J^{PC})$, and configurations for each state are given in columns 1 and 2. The numerical results of pure configuration and the configurations mixing, and the corresponding configurations mixing coefficients in Model I are presented in columns 3 to 5, while the calculation results of Model II are given in columns 6 to 8.}\label{nnnn states}
				\renewcommand
				\tabcolsep{0.08cm}
				\renewcommand{\arraystretch}{1.0}
				\begin{tabular}{ccccccccc}
					\hline\hline
					$I^{G}(J^{PC})$&Conf.&\multicolumn{3}{c}{\makecell{Model I}}& &\multicolumn{3}{c}{\makecell{Model II}}\\\cline{3-5}\cline{7-9}
					
				&&Pure&\multicolumn{2}{c}{\makecell{Configurations mixing}}&&Pure& \multicolumn{2}{c}{\makecell{Configurations mixing}}\\
					\cline{4-5}\cline{8-9}
					&&E(MeV)&E(MeV) &Mixing coefficients &&E(MeV)&E(MeV) &Mixing coefficients \\\hline
					$0^{+}(0^{++})$&$\vert1\rangle$&$1781.58$&$493.67$&$(-0.09,0.81,-0.12,0.55)$&&$1894.07$&$468.31$&$(0.16,0.78,-0.10,-0.59)$\\
					 &$\vert2\rangle$&$689.27$&$1077.63$&$(-0.18,-0.57,-0.16,0.78)$ & &$699.61$&$1063.19$&$(-0.25,0.62,0.10,0.73)$ \\&$\vert7\rangle$&$1813.58$&$1684.20$&$(0.74,-0.02,-0.67,0.02)$ & &$1831.20$&$1736.18$&$(0.56,0.03,0.82,0.05)$ \\&$\vert8\rangle$&$969.55$&$2002.63$&$(0.63,-0.01,0.71,0.29)$ & &$896.01$&$2126.75$&$(0.77,0.01,-0.55,0.32)$ \\\hline
					$0^{-}(1^{+-})$&$\vert7\rangle$&$1789.91$&$1116.17$&$(-0.21,0.98)$& &$1786.10$&$1116.52$&$(0.21,0.98)$\\&$\vert8\rangle$&$1148.6$&$1822.33$&$(0.98,0.21)$ & &$1149.14$&$1818.72$&$(0.98,-0.21)$ \\\hline
					$0^{+}(2^{++})$&$\vert7\rangle$&$1740.48$&$1423.99$&$(0.28,0.96)$&&$1680.86$&$1429.79$&$(-0.39,0.92)$\\&$\vert8\rangle$&$1450.27$&$1766.76$&$(0.96,-0.28)$ & &$1475.90$&$1726.97$&$(0.92,0.39)$ \\\hline
					$1^{-}(0^{++})$&$\vert1\rangle$&$1654.96$&$1654.96$&$(1.00,0.00)$&&$1719.87$&$1618.43$&$(0.00,1.00)$ \\
					&$\vert7\rangle$&$1656.67$&$1656.67$&$(0.00,1.00)$&&$1618.43$&$1719.87$&$(1.00,0.00)$ \\\hline
					$1^{-}(1^{++})$&$\vert3'\rangle$&$1256.53$&$1237.95$&$(0.97,0.24)$&&$1219.48$&$1199.41$&$(0.94,0.23)$\\
					 &$\vert5'\rangle$&$1534.87$&$1553.44$&$(0.24,0.97)$& &$1573.49$&$1593.56$&$(-0.23,0.94)$ \\\hline
					$1^{+}(1^{+-})$&$\vert4'\rangle$&$1109.68$&$1030.94$&$(0.91,0.41,0.00)$&&$1079.83$&$996.15$&$(0.91,0.41,0.00)$\\
					&$\vert6'\rangle$&$1401.55$&$1480.29$&$(-0.41,0.91,0.00)$&&$1419.55$&$1503.24$&$(-0.41,0.91,0.00)$ \\
					&$\vert7\rangle$&$1655.82$&$1655.82$&$(0.00,0.00,1.00)$&&$1616.35$&$1616.35$&$(0.00,0.00,1.00)$ \\\hline
					$1^{-}(2^{++})$&$\vert7\rangle$&$1654.10$&&&&$1612.13$&& \\\hline
					$2^{+}(0^{++})$&$\vert1\rangle$&$1284.86$&$665.17$&$(-0.62,0.78)$&&$1289.24$&$659.67$&$(-0.62,0.78)$ \\
					&$\vert7\rangle$&$1058.69$&$1678.39$&$(0.78,0.62)$&&$1061.64$&$1691.21$&$(0.78,0.62)$\\\hline
					$2^{-}(1^{+-})$&$\vert7\rangle$&$1210.20$&&& &$1185.61$&&\\\hline
					$2^{+}(2^{++})$&$\vert7\rangle$&$1469.36$&&&&$1411.61$&& \\
					\hline\hline
				\end{tabular}
			\end{table*}

\wuhao{\begin{table*}[htbp]
				\caption{A comparison of the calculated masses of $nn\bar{n}\bar{n}$ states for the observed exotic states. Columns 2 and 3 present the results from Model I and II. Columns 4 to 6 list the mass predictions from other representative studies of tetraquark interpretations.}\label{nnnn comparison}
				\renewcommand
				\tabcolsep{0.11cm}
				\renewcommand{\arraystretch}{1.0}
				\begin{tabular}{ccccccc}
					\hline\hline
					State&PDG~\cite{ParticleDataGroup:2024cfk}(MeV)&Model I(MeV)&Model II(MeV)&Ref.~\cite{Zhao:2021jss} (MeV)&Ref.~\cite{Deng_2012}(MeV)&Ref.~\cite{Ebert:2008id}(MeV)\\\hline\hline
	$f_0(500)$&$400\sim800$&494&468&$\diagup$&587&596\\
    $h_1(1170)$&1166&1116&1117&$\diagup$&1304&$\diagup$\\
    $f_2(1430)$&1440&1424&1430&$\diagup$&1465&$\diagup$\\
    $f_2(1640)$&1639&1767&1727&$\diagup$&1641&$\diagup$\\
    $a_1(1260)$&1230&1238&1199&$\diagup$&1839&1201\\
    $a_1(1640)$&1655&1553&1594&$\diagup$&2271&$\diagup$\\
    $a_2(1700)$&1706&1654&1612&$\diagup$&1807&$\diagup$\\
					\hline\hline
				\end{tabular}
			\end{table*}}

			For the $nn\bar{n}\bar{n}$ states, there are four energies predicted in the $0^+(0^{++})$ sector, where the lowest energy is responded to $f_{0}(500)$. $f_{0}(500)$ is observed in the low energy $\pi\pi$ scattering~\cite{Basdevant:1972uu} as well as in the $\pi\pi$ invariant mass distributions in $J/\psi\left(1S\right)\to\gamma\left(\pi\pi\right)$~\cite{Sarantsev:2021ein}. It has been considered as a candidate for scalar glueball~\cite{Mennessier:2010xg,Mennessier:2008kk,Kaminski:2009qg,Sarantsev:2021ein} and a dynamically generated pole in the $\pi\pi-K\bar{K}$ scattering \cite{Oller:1997ti}, where the average mass and width of $f_{0}(500)$ are $M_{f_{0}(500)} =400\sim800\,\mathrm{MeV},\,\Gamma_{f_{0}(500)} =100\sim800\,\mathrm{MeV}$ ~\cite{ParticleDataGroup:2024cfk}. In Model I, the lowest energy of $nn\bar{n}\bar{n}$ state with $I^G(J^{PC})=0^+(0^{++})$ is 493.67~MeV, and in Model II it is 468.31~MeV. The predictions of both models agree well with the physical mass of $f_{0}(500)$, suggesting that $f_{0}(500)$ may contain a sizable tetraquark component. Additionally, the highest energies predicted in both models are higher than the two-meson thresholds and may decay into $\eta\omega$, $\eta'\omega$, and $\eta'\eta'$.

            In the $I^G(J^{PC})=0^-(1^{+-})$ sector, Model I yields two states with the energies of 1116.17~MeV and 1822.33~MeV, while Model II yields 1116.52.19~MeV and 1818.72~MeV, respectively. The lower state is a plausible candidate for $h_1(1170)$, whose mass is $1166\pm5\pm3$~MeV~\cite{Dankowych:1981ks,Protopopescu:1973sh,Ando:1990ti}. Concerning $h_1(1170)$ being interpretable as $\left[q\bar{q}\right]$~\cite{Brau_1998}, the physical $h_1$ state might be a hybrid and decay into $\rho\pi$.

            In the $I^G(J^{PC})=0^+(2^{++})$ sector, Model I yields two states with energies of 1423.99~MeV and 1766.76~MeV, while Model II yields 1429.79~MeV and 1726.97~MeV, respectively. The BaBar collaboration reports the observation of the $f_2(1430)$ resonance in the processes $\gamma\gamma\to\eta_{c}(1S)\to\eta'\pi^{+}\pi^{-}$~\cite{BaBar:2021fkz}, where the mass and width of $f_2(1430)$ are $1440\pm11\pm3$~MeV and $46\pm15\pm5$~MeV, respectively. The lowest energy levels in the models match the $f_{2}(1430)$ mass. Moreover, the highest energies in these models correspond to $f_2(1640)$, whose mass is $1639\pm6$~$\rm{MeV}$~\cite{ParticleDataGroup:2024cfk,IHEP-IISN-LANL-LAPP-TSUIHEP:1988hke,IHEP-IISN-LANL-LAPP-TSUIHEP:1990rzz}. In Ref.~\cite{Deng_2012}, Deng $et\,al.$ employ a novel flux-tube structure to calculate the energy spectrum of tetraquark states $f_2(1640)$ is interpreted as a tetraquark with quantum numbers $N^{2S+1}L_{J}=2^1D_2$, as seen in Table~\ref{nnnn comparison}. Therefore, $f_2(1640)$ may contain both $S-$ and $D-$wave tetraquark components.

            In the $I^G(J^{PC})=1^-(0^{++})$ sector, both models predict two states within the region of $1600-1700$~MeV, suggesting the possible existence of $S-$wave tetraquark state with $I^G(J^{PC})=1^-(0^{++})$. To confirm this isovector scalar meson, further physical investigations are anticipated.

			In the $I^G(J^{PC})=1^-(1^{++})$ sector, Model I yields two states with energies of 1237.95~MeV and 1553.44~MeV, while Model II yields 1199.41~MeV and 1593.56~MeV, respectively. The COMPASS collaboration reports the comprehensive resonance-model fit of $\pi^{-}\pi^{-}\pi^{+}$ states.~\cite{COMPASS:2018uzl}, including the $a_{1}(1260)$ observed in the diffractive dissociation of negative pions into the $\pi^{-}\pi^{-}\pi^{+}$ state~\cite{COMPASS:2009xrl}, the $a_{1}(1420)$ observed in the diffractive dissociation of $190\,\mathrm{GeV}$ pions into the $\pi^{-}\pi^{-}\pi^{+}$ state~\cite{COMPASS:2015kdx}, and the $a_{1}(1640)$ observed in the reaction $\pi^{-} p\to\pi^{+}\pi^{-}\pi^{-} p$ state~\cite{Chung:2002pu}, where the mass and widths are
			 \begin{align*}
			 	(M_{a_{1}(1260)}=1230\pm40,\, & \Gamma_{a_{1}(1260)}=250\sim600)\,\mathrm{MeV},\\
			 	(M_{a_{1}(1420)}=1414^{+15}_{-13},\, & \Gamma_{a_{1}(1420)}=153^{+8}_{-23})\,\mathrm{MeV},\\
			 	(M_{a_{1}(1640)}=1655\pm16,\, & \Gamma_{a_{1}(1640)}=250\pm40)\,\mathrm{MeV}.\
			\end{align*}
			The lowest energy predictions from both models agree with $a_{1}(1260)$, and the highest energy predictions respond to $a_{1}(1640)$. In contrast, the mass of $a_{1}(1420)$ exhibits significant deviations from both model predictions, effectively excluding the interpretation as compact tetraquark states. In Ref.~\cite{Wang:2014bua}, the mass of the axial-vector tetraquark state with $I^G(J^{PC})=1^-(1^{++})$ is predicted to be approximately 1830~MeV in QCD sum rules. This result disfavors the interpretation of $a_{1}(1420)$ as a pure tetraquark state, which is consistent with the predictions in our models.

            In the $I^G(J^{PC})=1^-(2^{++})$ sector, Model I yields a state with energy of 1654.10~MeV, and Model II gives 1612.13~MeV. These values suggest that this state is a plausible candidate for $a_2(1700)$, whose mass is $1706\pm14$ MeV~\cite{ParticleDataGroup:2024cfk,L3:1997mpi,Kopf:2020yoa}. Concerning $a_2(1700)$ being interpretable as $\left[ q\bar{q}\right]$~\cite{COMPASS:2018uzl}, the physical $a_2$ state might be a hybrid and decay into $\pi\pi$.
           In addition, several isotensor states are also displayed; see Table~\ref{nnnn states}.
	
			\subsection{$ns\bar{n}\bar{s}$ and $ss\bar{s}\bar{s}$ Systems.}
\begin{figure}[htbp]
				\centering
				\includegraphics[width=1\textwidth]{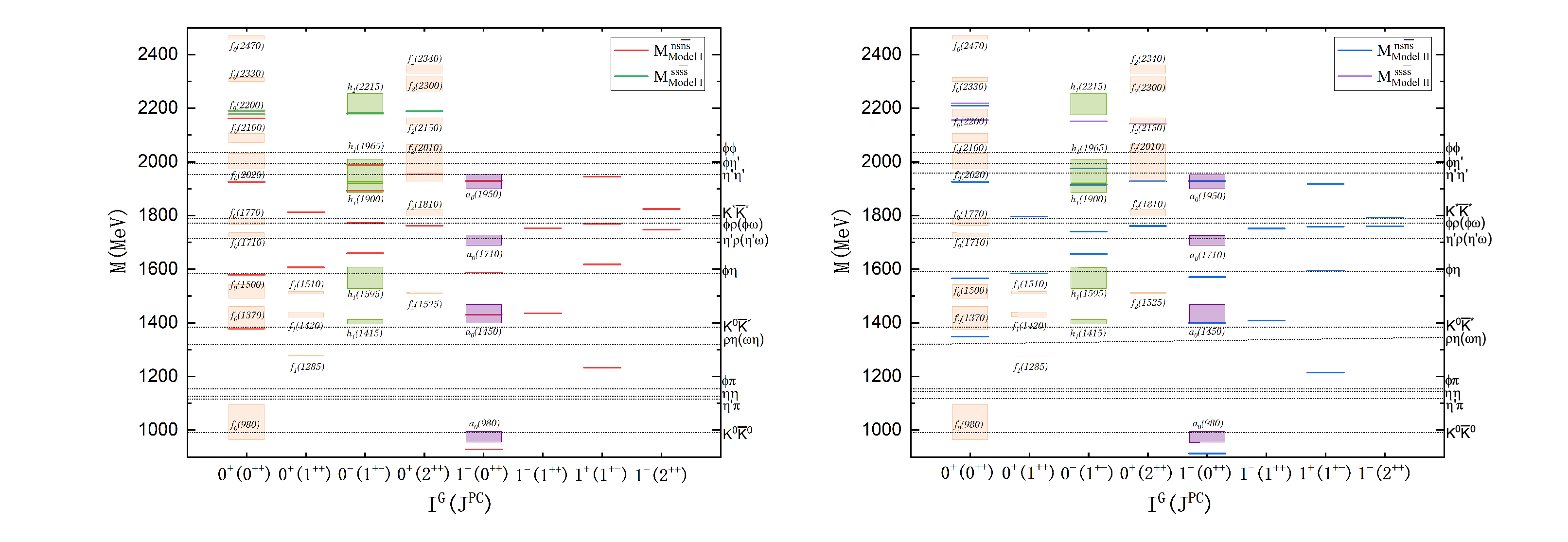}
				\caption{Mass spectrum of S-wave $ns\bar{n}\bar{s}$ and $ss\bar{s}\bar{s}$ states. The data on the left corresponds to Model I, whereas the data on the right pertains to Model II. The red and green solid lines depict the numerical results of the $ns\bar{n}\bar{s}$ and $ss\bar{s}\bar{s}$ states, respectively, as calculated within Model I. Conversely, the blue and purple lines represent the numerical results of these states as determined by Model II. The gray dotted lines are the two-body thresholds, the rectangles represent $ns\bar{n}\bar{s}$ and $ss\bar{s}\bar{s}$ states' physical masses quoted from PDG~\cite{ParticleDataGroup:2024cfk}, and the bands of the rectangles stand for the uncertainties of masses.} \label{nsns picture}
			\end{figure}

            The configurations of $ns\bar{n}\bar{s}$ and $ss\bar{s}\bar{s}$ are organized in the Table~\ref{Configurations of nsns Systems}, where the energy eigenstates are presented in Table~\ref{nsns states}. Additionally, the mass spectra of $ns\bar{n}\bar{s}$ and $ss\bar{s}\bar{s}$ are displayed in Fig.~\ref{nsns picture}.
             \begin{table}[htbp]
				\caption{Configurations of $ns\bar{n}\bar{s}$ and $ss\bar{s}\bar{s}$ Systems.}\label{Configurations of nsns Systems}
				\renewcommand
				\tabcolsep{0.72cm}
				\renewcommand{\arraystretch}{1.0}
				\begin{tabular}{cccccc}
					\toprule
					\multicolumn{6}{c}{For $ns\bar{n}\bar{s}$ states}\\\hline
					$I^{G}$&$J^{PC}$&\multicolumn{4}{c}{Configurations}\\\hline
					$0^{+}$&$0^{++}$&$|1\rangle$&$|2\rangle$&$|7\rangle$&$|8\rangle$\\\hline
					$0^{+}$&$1^{++}$&$|3'\rangle=\frac{1}{\sqrt{2}}\left(|3\rangle+|4\rangle\right)$ &  $|5'\rangle=\frac{1}{\sqrt{2}}\left(|5\rangle+|6\rangle\right)$ &&\\\hline
					$0^{-}$&$1^{+-}$& $|4'\rangle=\frac{1}{\sqrt{2}}\left(|3\rangle-|4\rangle\right)$ &   $|6'\rangle=\frac{1}{\sqrt{2}}\left(|5\rangle-|6\rangle\right)$ &  $|7\rangle$  &   $|8\rangle$ \\\hline
					$0^{+}$&$2^{++}$ &$|7\rangle$ &   $|8\rangle$ &&\\\hline		
					$1^{-}$&$0^{++}$&$|1\rangle$&$|2\rangle$ &$|7\rangle$&$|8\rangle$\\\hline
					$1^{-}$&$1^{++}$&$|3'\rangle=\frac{1}{\sqrt{2}}\left(|3\rangle+|4\rangle\right)$ &  $|5'\rangle=\frac{1}{\sqrt{2}}\left(|5\rangle+|6\rangle\right)$ &&\\\hline
					$1^{+}$&$1^{+-}$& $|4'\rangle=\frac{1}{\sqrt{2}}\left(|3\rangle-|4\rangle\right)$ &   $|6'\rangle=\frac{1}{\sqrt{2}}\left(|5\rangle-|6\rangle\right)$ &  $|7\rangle$  &   $|8\rangle$ \\\hline
					$1^{-}$&$2^{++}$ &$|7\rangle$ &   $|8\rangle$ &&\\\hline		
					\multicolumn{6}{c}{For $ss\bar{s}\bar{s}$ states}\\\hline
					$I^{G}$&$J^{PC}$&\multicolumn{4}{c}{Configurations}\\\hline
					$0^{+}$ &$0^{++}$& $|1\rangle$&$|7\rangle$&&\\\hline
					$0^{-}$&$1^{+-}$&  $|7\rangle$ &&&\\\hline
					$0^{+}$&$2^{++}$&$|7\rangle$&&&\\
					\toprule
				\end{tabular}
			\end{table}

            \begin{table*}[hbtp]
				\caption{Numerical results for the $ns\bar{n}\bar{s}$ and $ss\bar{s}\bar{s}$ states in two models. The quantum numbers $I^{G}(J^{PC})$, and configurations for each state are given in columns 1 and 2. The numerical results of pure configuration and the configurations mixing, and the corresponding configurations mixing coefficients in Model I are presented in columns 3 to 5, while the calculation results of Model II are given in columns 6 to 8.}\label{nsns states}
				\renewcommand
				\tabcolsep{0.08cm}
				\renewcommand{\arraystretch}{1.0}
				\begin{tabular}{ccccccccc}
					\hline\hline

					$I^{G}(J^{PC})$&Conf.&\multicolumn{3}{c}{\makecell{Model I}}& &\multicolumn{3}{c}{\makecell{Model II}}\\\cline{3-5}\cline{7-9}
					
					&&Pure&\multicolumn{2}{c}{\makecell{Configueations mixing}}&&Pure& \multicolumn{2}{c}{\makecell{Configueations mixing}}\\
					\cline{4-5}\cline{8-9}
					&&E(MeV)&E(MeV) &Mixing coefficients &&E(MeV)&E(MeV) &Mixing coefficients \\\hline
					\multicolumn{9}{c}{\makecell{For $ns\bar{n}\bar{s}$ state}}\\\hline
                    $0^{+}(0^{++})$&$\vert1\rangle$&$1894.70$&$1378.43$&$(0.35,0.50,-0.46,0.64)$&&$1934.56$&$1349.06$&$(-0.36,-0.53,0.46,0.61)$\\
					  &$\vert2\rangle$&$1536.21$&$1579.18$&$(-0.25,0.85,0.13,-0.43)$& &$1510.52$&$1565.37$&$(-0.25,0.84,0.18,0.44)$\\
					&$\vert7\rangle$&$1883.92$&$1925.20$&$(0.74,0.10,0.65,-0.02)$ & &$1846.75$&$1924.92$&$(0.65,0.09,0.74,-0.10)$ \\
					 &$\vert8\rangle$&$1729.54$&$2161.56$&$(-0.50,0.08,0.58,0.63)$& &$1757.15$&$2209.63$&$(0.61,-0.06,-0.45,0.65)$ \\\hline
					$0^{+}(1^{++})$&$\vert3'\rangle$&$1730.19$&$1606.57$&$(-0.63,0.77)$&&$1685.11$&$1584.10$&$(0.72,0.69)$\\
					&$\vert5'\rangle$&$1688.66$&$1812.29$&$(0.77,0.63)$& &$1694.91$&$1795.92$&$(-0.69,0.72)$ \\\hline
					$0^{-}(1^{+-})$&$\vert4'\rangle$&$1800.60$&$1659.82$&$(-0.09,-0.02,-0.53,0.84)$&&$1757.73$&$1656.33$&$(0.16,-0.03,0.61,0.77)$\\
				 &$\vert6'\rangle$&$1862.86$&$1771.54$&$(0.87,-0.49,0.00,0.00)$&&$1896.99$&$1740.27$&$(0.95,-0.31,0.00,0.00)$ \\
				  &$\vert7\rangle$&$1893.28$&$1891.92$&$(-0.49,0.87,0.00,0.00)$&&$1852.63$&$1914.45$&$(0.31,0.95,0.00,0.00)$ \\
				  &$\vert8\rangle$&$1754.90$&$1988.36$&$(-0.11,0.05,0.83,0.53)$&&$1779.50$&$1975.80$&$(-0.10,-0.06,0.78,0.61)$\\\hline
				$0^{+}(2^{++})$&$\vert7\rangle$&$1911.59$&$1761.97$&$(0.47,0.88)$&&$1864.34$&$1760.91$&$(-0.61,0.79)$ \\
					 &$\vert8\rangle$&$1804.14$&$1953.75$&$(0.88,-0.46)$&&$1823.96$&$1927.39$&$(0.79,0.61)$ \\\hline
				$1^{-}(0^{++})$&$\vert1\rangle$&$1725.26$&$927.46$&$(-0.07,0.64,-0.09,0.76)$&&$1737.59$&$912.70$&$(-0.08,-0.62,0.11,0.77)$\\
					&$\vert2\rangle$&$1310.28$&$1430.18$&$(0.62,0.22,0.74,-0.04)$& &$1303.68$&$1399.10$&$(-0.58,0.23,0.78,0.01)$\\
					&$\vert7\rangle$&$1632.85$&$1588.32$&$(-0.10,0.73,-0.17,0.64)$ & &$1586.33$&$1570.22$&$(0.09,0.74,-0.16,0.64)$ \\
					&$\vert8\rangle$&$1206.58$&$1929.01$&$(0.77,-0.02,-0.63,0.01)$& &$1182.98$&$1928.55$&$(0.80,0.02,0.60,0.02)$ \\\hline
					$1^{-}(1^{++})$&$\vert3'\rangle$&$1554.51$&$1434.78$&$(0.79,-0.61)$&&$1522.43$&$1407.96$&$(0.81,0.58)$\\
				 &$\vert5'\rangle$&$1632.67$&$1752.40$&$(0.61,0.79)$& &$1637.14$&$1751.62$&$(-0.58,0.81)$ \\\hline
                 $1^{+}(1^{+-})$&$\vert4'\rangle$&$1576.82$&$1232.35$&$(-0.49,-0.42,-0.07,0.75)$&&$1544.54$&$1214.53$&$(0.50,-0.42,0.08,0.75)$\\&$\vert6'\rangle$&$1684.54$&$1617.34$&$(0.62,0.35,0.27,0.64)$&&$1693.79$&$1595.54$&$(0.67,-0.26,0.28,-0.63)$ \\
				 &$\vert7\rangle$&$1893.78$&$1768.38$&$(-0.60,0.72,0.35,0.04)$ & &$1851.94$&$1758.10$&$(0.53,0.69,-0.48,0.08)$ \\
				&$\vert8\rangle$&$1407.08$&$1944.14$&$(0.00,-0.43,0.89,-0.15)$& &$1394.96$&$1917.06$&$(0.03,0.52,0.83,0.19)$ \\\hline
				$1^{-}(2^{++})$&$\vert7\rangle$&$1819.26$&$1747.17$&$(0.25,0.97)$&&$1777.91$&$1759.23$&$(-0.65,0.76)$ \\
				  &$\vert8\rangle$&$1752.08$&$1824.17$&$(0.97,-0.25)$&&$1773.25$&$1791.92$&$(0.76,0.65)$ \\\hline
					\multicolumn{9}{c}{\makecell{For $ss\bar{s}\bar{s}$ state}}\\\hline
                    $0^{+}(0^{++})$&$\vert1\rangle$&$2189.93$&$2177.52$&$(0.00,1.00)$&&$2217.96$&$2155.79$&$(0.00,1.00)$ \\
					&$\vert7\rangle$&$2177.52$&$2189.93$&$(1.00,0.00)$&&$2155.79$&$2217.96$&$(1.00,0.00)$\\\hline

					$0^{-}(1^{+-})$&$\vert7\rangle$&$2181.24$&&& &$2151.32$&&\\\hline
				$0^{+}(2^{++})$&$\vert7\rangle$&$2188.25$&&&&$2141.86$&& \\
					\hline\hline
				\end{tabular}
			\end{table*}

  \wuhao{      \begin{table*}[htbp]
				\caption{A comparison of the calculated masses of $ns\bar{n}\bar{s}$ and $ss\bar{s}\bar{s}$ states for the observed exotic states. Columns 2 and 3 present the results from Model I and II. Columns 4 to 6 list the mass predictions from other representative studies of tetraquark interpretations.}\label{nsns comparison}
				\renewcommand
				\tabcolsep{0.11cm}
				\renewcommand{\arraystretch}{1.0}
				\begin{tabular}{ccccccc}
					\hline\hline
					State&PDG~\cite{ParticleDataGroup:2024cfk}(MeV)&Model I(MeV)&Model II(MeV)&Ref.~\cite{Zhao:2021jss} (MeV)&Ref.~\cite{Deng_2012}(MeV)&Ref.~\cite{Ebert:2008id}(MeV)\\\hline\hline
	$f_0(1370)$&1422&1378&1349&1431&1318&1179\\
    $f_0(2020)$&1982&1925&1925&1986&1925&$\diagup$\\
    $f_0(2200)$&2178&2162&2210&2252&2095&2203\\
    $h_1(1595)$&1594&1660&1656&1678&1586&$\diagup$\\
    $h_1(1900)$&1911&1892&1914&$\diagup$&$\diagup$&$\diagup$\\
    $h_1(1965)$&1965&1988&1976&2081&$\diagup$&1942\\
    $f_2(1810)$&1815&1762&1761&$\diagup$&1755&$\diagup$\\
    $f_2(2010)$&2011&1954&1927&$\diagup$&1946&2097\\
    $a_0(980)$&980&927&913&$\diagup$&1318&992\\
    $a_0(1450)$&1439&1430&1399&$\diagup$&1590&1480\\
    $a_0(1950)$&1931&1929&1929&$\diagup$&$\diagup$&$\diagup$\\
    $h_1(2215)$&2215&2181&2151&$\diagup$&$\diagup$&2267\\
    $f_2(2150)$&2157&2188&2141&$\diagup$&$\diagup$&$\diagup$\\
					\hline\hline
				\end{tabular}
			\end{table*}}
			
			For the $ns\bar{n}\bar{s}$ tetraquark states in the $0^+(0^{++})$ sector, four distinct mass predictions are obtained. Among these, the lowest masses were calculated to be 1378.43~MeV and 1349.06~MeV, which align closely with $f_{0}(1370)$, rather than $f_{0}(980)$. Thus, $f_{0}(980)$ cannot be described as a compact tetraquark state in the present models. Due to the masses of $f_{0}(980)$ being close to the $K\bar{K}$ threshold, the $K\bar{K}$ channel plays a crucial role in this region~\cite{Zou:1993az,Baru:2003qq}. In Ref.~\cite{Branz:2007xp}, the decay widths for both the process $f_{0}(980)\to\gamma\gamma$ and the interaction decay $f_{0}(980)\to\pi\pi$ are calculated using the phenomenological Lagrangian approach. The results provide strong evidence that $f_{0}(980)$ is mainly the hadronic molecular state of $K\bar{K}$.

            $f_{0}(1370)$ is observed in the $\pi^+\pi^+\pi^-\pi^-$ invariant mass distributions in $\bar{p}n\to\pi^+\pi^+\pi^-\pi^-\pi^-$~\cite{Bettini:1966zz} as well as in the $K\bar{K}$ invariant mass distributions in $D^{+}\to K^{-}K^{+}K^{+}$~\cite{LHCb:2019tdw}. The mass and width of $f_{0}(1370)$ are found to be $1422\pm15\pm9\pm28$~MeV and $324\pm38\pm18\pm38$~MeV. The lowest energy predictions align with the mass of $f_{0}(1370)$. Furthermore, the second-lowest energy predictions respond to $f_{0}(1500)$, whose mass and width are $1522\pm25$~MeV and $108\pm33$~MeV~\cite{ParticleDataGroup:2024cfk,Gray:1983cw,Serpukhov-Brussels-AnnecyLAPP:1983ydj,BESIII:2022sfx,BESIII:2022iwi}. In Ref.~\cite{Li:2000yn}, Li $et\,al.$ study the two-hadronic decays of $f_{0}(1370)$, $f_{0}(1500)$, $f_{0}(1710)$ in the admixture of the scalar glueball and quarkonia, where $f_{0}(1370)$, $f_{0}(1500)$, and $f_{0}(1710)$ are strong candidates for the glueball states. Therefore, $f_{0}(1370)$ and $f_{0}(1500)$ might be hybrid states with a sizable tetraquark component.

			$f_{0}(2020)$ is observed in the invariant mass of the $\eta'\eta'$ system in $J/\psi\to\gamma\eta'\eta'$~\cite{WA102:1997sum,BESIII:2022zel}, where the mass and width are $1982\pm3^{+54}_{-0}$ MeV and $436\pm4^{+46}_{-49}$~MeV, respectively. The second-highest energy predictions of $ns\bar{n}\bar{s}$ state with $I^G(J^{PC})=0^+(0^{++})$ are 1925.20~MeV and 1924.92~MeV, which are consistent with the physical mass of $f_{0}(2020)$. Thus, $f_{0}(2020)$ may contain a sizable tetraquark component. On the other hand,
            $f_{0}(2200)$ is reported in the $\eta\eta$ invariant mass from $\pi^{-}p$ charge-exchange reactions at $32.5\,\mathrm{GeV/c}$~\cite{Sarantsev:2021ein}, where the mass and width are $M_{f_{0}(2200)} =2187\pm14\,\mathrm{MeV},\,\Gamma_{f_{0}(2200)} =210\pm40\,\mathrm{MeV}$. In Model I, the lowest energy of $ss\bar{s}\bar{s}$ state with $I^G(J^{PC})=0^+(0^{++})$ is 2192.51~MeV, which is very consistent with the mass of $f_{0}(2200)$ state. The predictions with the highest energy of $ns\bar{n}\bar{s}$ state with $I^G(J^{PC})=0^+(0^{++})$ are 2161.56~MeV and 2209.63~MeV, which are consistent with the physical mass of $f_{0}(2200)$. Thus, $f_{0}(2200)$ may contain a sizable tetraquark component. As shown in Table~\ref{nsns comparison}, Ref.~\cite{Zhao:2021jss} predicts that the $2S$-wave light tetraquark states with $I=0,\,J=0$ calculated within nonrelativistic quark models, are associated with $f_{0}(2020)$ and $f_{0}(2200)$. Therefore, $f_{0}(2020)$ and $f_{0}(2200)$ may contain a hybrid structure both in the $1S$- and $2S$- isoscalar scalar mesons.

            For the $ns\bar{n}\bar{s}$ system, four different energy levels are predicted in the $I^G(J^{PC})=0^-(1^{+-})$ sector. The lowest energy predictions are 1659.82~MeV and 1656.33~MeV, respond to $h_1(1595)$, whose mass is $1594\pm15^{+10}_{-60}$ MeV~\cite{BNL-E852:2000poa}. The second-highest energy predictions are 1891.92~MeV and 1914.45~MeV, matching with the $h_{1}(1900)$ structure in the $\phi\eta$ invariant mass in $J/\psi\to\phi\pi^0\eta$~\cite{BESIII:2023zwx}, where the mass and width are $1911\pm6(stat.)\pm14(sys.)$~MeV and $149\pm12(stat.)\pm23(sys.)$~MeV. The predictions with the highest energy are 1988.36~MeV and 1975.80~MeV. These values suggest that this state is a plausible candidate for $h_1(1965)$, whose mass and width are $1965\pm45$~MeV and $345\pm75$~\cite{Anisovich:2002xoo}. In Ref.~\cite{Zhao:2021jss}, the $1S$-wave light tetraquark state with $I=0,\,J=1$ predicted by nonrelativistic quark models align with the $h_{1}(1595)$, while the $2S$-wave state correspond to $h_{1}(1965)$. Further analysis by Wang $et\,al.$ using the modified Godfrey$-$Isgur model and QPC approach suggests that $h_1(1595)$ is a promising candidate for the $2^1P_1$ $n\bar{n}$ state, $h_1(1900)$ is more likely to correspond to the $2^1P_1$ $s\bar{s}$, and $h_1(1965)$ may correspond to the $3^1P_1$ $q\bar{q}$ state~\cite{Wang:2024lba}. These interpretations imply that these states may have hybrid structures: $h_1(1595)$ could a mixture of a $2P$- $n\bar{n}$ and a $1S$-$ns\bar{n}\bar{s}$ tetraquark component; $h_1(1900)$ might incorporate both a $2P$-wave $s\bar{s}$ component and a $1S$-wave $ns\bar{n}\bar{s}$ tetraquark component; while $h_1(1965)$ may feature a $3P$-wave $n\bar{n}$ component mixed with both $1S$-wave and $2S$-wave $ns\bar{n}\bar{s}$ tetraquark components.

            For the $I^G(J^{PC})=0^+(2^{++})$ sector, there are two energy states predicted in both models. The lowest energy predictions are 1761.97~MeV and 1760.91~MeV. These values suggest that this state is a plausible candidate for $f_2(1810)$, whose mass is $1815\pm12$~MeV~\cite{ParticleDataGroup:2024cfk,BARI-BONN-CERN-GLASGOW-LIVERPOOL-MILAN-VIENNA:1980wqx,BESIII:2013qqz}. Concerning $f_2(1810)$ being interpretable as $\left[q\bar{q}\right]$~\cite{BESIII:2013qqz}, the physical $f_2$ state might be a hybrid and decay into $\eta\eta$.
            In Refs.~\cite{Etkin:1985se,Etkin:1987rj}, Etkin $et\,al.$ perform a partial wave analysis of the reaction $\pi^-p\to\phi\phi n$, and reveal three resonances with quantum numbers $I^G(J^{PC})=0^+(2^{++})$ in the $\phi\phi$ channel, named $g_{T}, g_{T'}, g_{T''}$. Later, they were named $f_{2}(2010), f_{2}(2300), f_{2}(2340)$, respectively, where the masses and widths are
			\begin{align*}
			(M_{f_{2}(2010)}=2011^{+62}_{-76},\, & \Gamma_{f_{2}(2010)}=202^{+67}_{-62})\,\mathrm{MeV},\\
			(M_{f_{2}(2300)}=2297\pm28,\,&\Gamma_{f_{2}(2300)}=149\pm41)\,\mathrm{MeV},\\
			(M_{f_{2}(2340)}=2339\pm55,\, & \Gamma_{f_{2}(2340)}=319^{+81}_{-69})\,\mathrm{MeV}.\
			\end{align*}
			 In Model I, the highest energy of $S-$wave $ns\bar{n}\bar{s}$ state with $I^G(J^{PC})=0^+(2^{++})$ is 1953.75~MeV, while in Model II it is 1927.39~MeV. Numerically, both are consistent with the mass of $f_{2}(2010)$, suggesting that $f_{2}(2010)$ may contain a significant $ns\bar{n}\bar{s}$ tetraquark state. In contrast, $f_{2}(2300)$ and $f_{2}(2340)$ are inconsistent with the predictions of two models and cannot be explained as a compact tetraquark state here.

             For the $I^G(J^{PC})=1^-(0^{++})$ sector, Model I yields four states with energies of 927.46~MeV, 1430.18~MeV, 1588.32~MeV, and 1929.01~MeV, while Model II yields 912.70~MeV, 1399.10~MeV, 1570.22~MeV, and 1928.55~MeV, respectively. The lowest energy predictions respond to $a_0(980)$, whose mass and width are $980\pm20\,\rm{MeV}$ and $50\sim100\,\rm{MeV}$~\cite{ParticleDataGroup:2024cfk,Astier:1967zz,BESIII:2022npc}. The second-lowest energy predictions respond to $a_0(1450)$, whose mass and width are $1439\pm34$~MeV and $258\pm14$~MeV~\cite{ParticleDataGroup:2024cfk,Cason:1976fn,LHCb:2015lnk,CrystalBarrel:2019zqh}. The predictions with the highest energy respond to $a_0(1950)$, whose mass and width are $1931\pm26$~MeV and $270\pm40$~MeV~\cite{ParticleDataGroup:2024cfk,BaBar:2015kii}. In Ref.~\cite{Deng_2012}, the $1S$ light tetraquark states with $I^G(J^{PC})=1^-(0^{++})$ correspond to $a_{0}(980)$ and the $2S$-wave light tetraquark states with $I^G(J^{PC})=1^-(0^{++})$ correspond to $a_{0}(1450)$. Therefore, the $a_{0}(980)$ state may contain a component of $1S$ tetraquark , while the $a_{0}(1450)$ is complicated.

Note that for the $ns\bar{n}\bar{s}$ tetraquark system, the present two models consistently predict the state with quantum numbers $I=0$ to be higher in energy than the corresponding $I=1$ state. This mass inversion results from the unique behavior of instanton-induced potentials in low-lying tetraquark configurations.
            The observation reveals:
            \begin{itemize}
            \item Conventional interactions maintain isospin symmetry, contributing equally to all states;
            \item The instanton potential exhibits a striking isospin dependence:
            \begin{itemize}
            \item For the interaction between $ns$ or $\bar{n}\bar{s}$: identical negative contributions to both $I=0$ and $I=1$ channels
            \item For the interaction between $n\bar{n}$:
            \begin{itemize}
            \item Positive energy shift ($+\Delta E$) in $I=0$ sector
            \item Negative shift ($-\Delta E$) in $I=1$ sector
            \end{itemize}
            \end{itemize}
            \end{itemize}
            This discrepancy pattern ultimately leads to the energy splitting between isospin eigenstates. To cancel the energy gap induced by instanton effects, a viable approach is to introduce $q\bar{q}$ mixing coupling, as detailed in Ref.~\cite{tHooft:2008rus}.

For the $ss\bar{s}\bar{s}$ tetraquark states, both models predict a state in the $0^-(1^{+-})$ sector and a state in the $0^+(2^{++})$ sector. The predictions in the $0^-(1^{+-})$ sector are 2181.24~MeV and 2151.32~MeV. These values suggest that this state is a plausible candidate for $h_1(2215)$, whose mass and width are $2215\pm40$~MeV and $325\pm55$~\cite{Anisovich:2002xoo}. The predictions in the $0^+(2^{++})$ sector are 2188.25~MeV and 2141.86~MeV. These values suggest that this state is a plausible candidate for $f_2(2150)$, whose mass and width are $2157\pm12$~MeV and $152\pm30$ MeV~\cite{ParticleDataGroup:2024cfk,Alspector:1973er,Cutts:1974vi,Uman:2006xb,CLEO:2010fre}. In Ref.~\cite{Ebert:2008id}, Ebert $et\,al.$ employ the quasipotential approach within quantum chromodynamics to evaluate the spectrum of light tetraquarks. They predicted the mass of the $ss\bar{s}\bar{s}$ state with $J^{PC}=1^{+-}$ is 2267~MeV, aligning with $h_1(2215)$. Additionally, there is a prediction the $ns\bar{n}\bar{s}$ state with $J^{PC}=2^{++}$ at 2097~MeV  which corresponds to $f_2(2150)$ \cite{Ebert:2008id}. Therefore, $h_1(2215)$ may contain a component of $ss\bar{s}\bar{s}$, while $f_2(2150)$ may contain components of $ns\bar{n}\bar{s}$ and $ss\bar{s}\bar{s}$.

			\subsection{$nn\bar{n}\bar{s}$ and $ns\bar{s}\bar{s}$ Systems.}
\begin{figure*}[hbtp]
				\centering
				\includegraphics[width=1\textwidth]{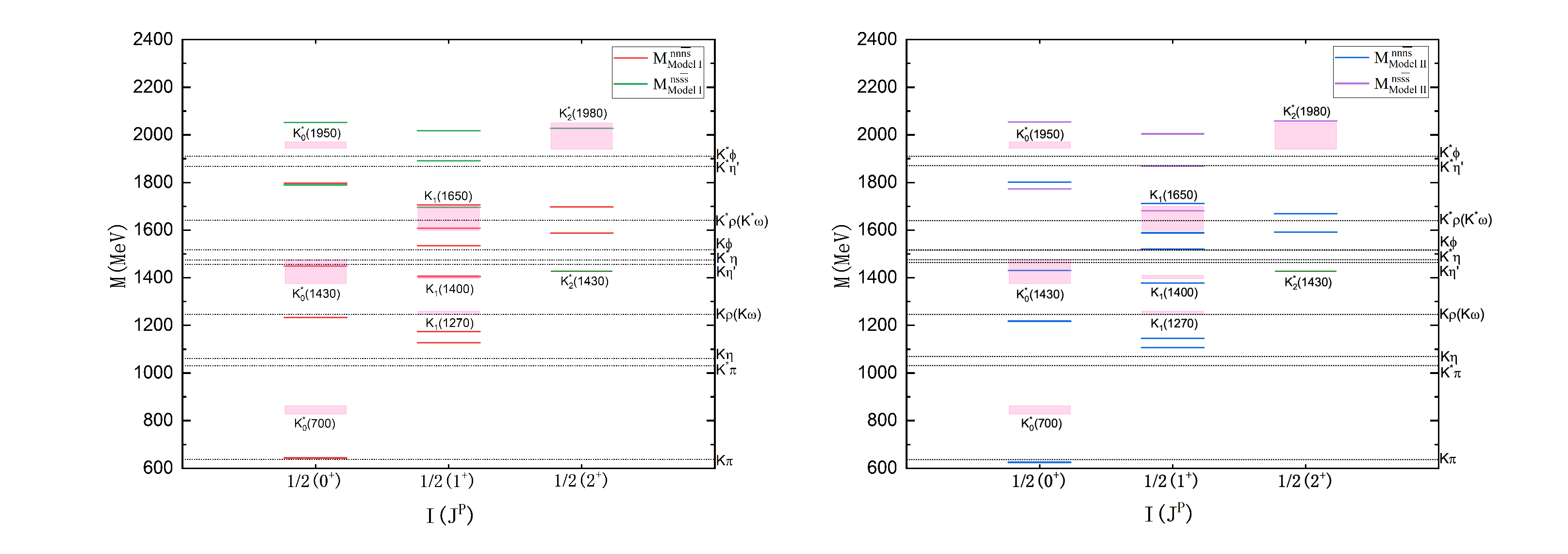}
				\caption{Mass spectrum of S-wave $nn\bar{n}\bar{s}$ and $ns\bar{s}\bar{s}$ states. The data on the left corresponds to Model I, whereas the data on the right pertains to Model II. The red and green solid lines depict the numerical results of the $nn\bar{n}\bar{s}$ and $ns\bar{s}\bar{s}$ states, respectively, as evaluated in Model I. Conversely, the blue and purple lines represent the corresponding outcomes for these states as determined by Model II. The gray dotted lines are the two-body thresholds, the rectangles represent $nn\bar{n}\bar{s}$ and $ns\bar{s}\bar{s}$ states physical masses taken from PDG \cite{ParticleDataGroup:2024cfk}, and the bands of the rectangles stand for the uncertainties of masses. }\label{nnns picture}
			\end{figure*}

            The configurations of $nn\bar{n}\bar{s}$ and $ns\bar{s}\bar{s}$ are organized in the Table~\ref{Configurations of nnns Systems}. The energy eigenstates are presented in Table~\ref{nnns states}, where the component of the tetraquark state, the corresponding quantum numbers $I(J^{P})$, and the configurations for each state are given in columns 1 and 3, the pure configuration calculation results, the configurations mixing results, and the corresponding configurations mixing coefficients in Model I are presented in columns 4 to 6, while the calculation results of Model II are given in columns 7 to 9. Additionally, the mass spectra of $nn\bar{n}\bar{s}$ and $ns\bar{s}\bar{s}$ are shown in Fig.~\ref{nnns picture}.
             \begin{table}[h]
				\caption{Configurations of $nn\bar{n}\bar{s}$ and $ns\bar{s}\bar{s}$ Systems.}\label{Configurations of nnns Systems}
				\renewcommand
				\tabcolsep{0.82cm}
				\renewcommand{\arraystretch}{1.0}
				\begin{tabular}{cccccccc}
					\toprule
					\multicolumn{8}{c}{ $nn\bar{n}\bar{s}$ states}\\\hline
					$I$&$J^{P}$&\multicolumn{6}{c}{Configurations}\\\hline
					$\frac{1}{2}$&$0^{+}$&$|1\rangle$&$|2\rangle$&$|7\rangle$&$|8\rangle$ &\\\hline
					$\frac{1}{2}$&$1^{+}$&$|3\rangle$&$|4\rangle$&$|5\rangle$&$|6\rangle$ &$|7\rangle$&$|8\rangle$ \\\hline
					$\frac{1}{2}$&$2^{+}$ &$|7\rangle$ &   $|8\rangle$&&&& \\\hline		
					$\frac{3}{2}$&$0^{+}$&$|1\rangle$&$|7\rangle$ &&&&\\\hline
					$\frac{3}{2}$&$1^{+}$&$|3\rangle$ &$|6\rangle$ &$|7\rangle$ &&&\\\hline
					$\frac{3}{2}$&$2^{+}$& $|7\rangle$ &&&&&\\\hline
					\multicolumn{8}{c}{For $ns\bar{s}\bar{s}$ states}\\\hline
					$I$&$J^{P}$&\multicolumn{6}{c}{Configurations}\\\hline
					$\frac{1}{2}$ &$0^{+}$& $|1\rangle$&$|7\rangle$&&&&\\\hline
					$\frac{1}{2}$&$1^{+}$&  $|4\rangle$&$|5\rangle$&$|7\rangle$ &&&\\\hline
					$\frac{1}{2}$&$2^{+}$&  $|7\rangle$&&&&&\\
					\toprule
				\end{tabular}
			\end{table}

\begin{sidewaystable}
            \centering
				\caption{Numerical results for the $nn\bar{n}\bar{s}$ and $ns\bar{s}\bar{s}$ states in two models.}\label{nnns states}
				\renewcommand
				\tabcolsep{0.16cm}
				\renewcommand{\arraystretch}{1.0}
				\begin{longtable}{ccccccccccc}
					\hline\hline

					State&$I(J^{P})$&Conf.&\multicolumn{3}{c}{\makecell{Model I}}& &\multicolumn{3}{c}{\makecell{Model II}}\\\cline{4-6}\cline{8-10}
					
					 &&&Pure&\multicolumn{2}{c}{\makecell{Configueations mixing}}&&Pure& \multicolumn{2}{c}{\makecell{Configueations mixing}}\\
					\cline{5-6}\cline{9-10}
					&&&E(MeV)&E(MeV) &Mixing coefficients &&E(MeV)&E(MeV) &Mixing coefficients \\\hline
					\endhead

                   $nn\bar{n}\bar{s}$& $\frac{1}{2}(0^{+})$&$\vert1\rangle$&$1593.33$&$643.25$&$(-0.16,0.68,0.21,0.68)$&&$1609.33$&$624.73$&$(0.17,0.66,0.22,0.69)$\\
				&&$\vert2\rangle$&$991.05$&$1233.22$&$(0.44,0.64,-0.51,-0.37)$& &$993.30$&$1217.90$&$(-0.44,0.63,-0.56,-0.31)$\\
				&&$\vert7\rangle$&$1493.91$&$1448.06$&$(0.42,-0.36,-0.54,0.63)$ & &$1448.73$&$1429.79$&$(-0.33,-0.41,-0.54,0.65)$ \\
				&&$\vert8\rangle$&$1040.78$&$1794.66$&$(0.77,-0.03,0.63,0.01)$& &$1022.79$&$1801.74$&$(0.81,0.03,-0.58,-0.05)$ \\\hline
                $nn\bar{n}\bar{s}$&$\frac{1}{2}(1^{+})$&$\vert3\rangle$&$1430.13$&$1127.41$&$(-0.46,-0.11,0.10,0.39,0.17,0.77)$&&$1391.39$&$106.76$&$(0.49,-0.16,-0.12,-0.40,-0.14,0.73)$\\
                &&$\vert4\rangle$&$1286.03$&$1173.66$&$(0.07,0.82,-0.47,-0.08,-0.12,0.29)$& &$1259.77$&$1145.81$&$(-0.04,0.82,-0.45,-0.04,-0.11,-0.33)$ \\
                &&$\vert5\rangle$&$1503.15$&$1404.91$&$(0.61,-0.17,0.04,-0.41,0.46,0.46)$&&$1510.89$&$1377.45$&$(0.63,0.14,0.05,-0.33,0.50,-0.47)$ \\&&$\vert6\rangle$&$1526.85$&$1533.85$&$(0.02,0.53,0.66,0.21,0.46,-0.18)$&&$1536.47$&$1520.11$&$(-0.14,-0.50,0.53,0.13,0.60,0.26)$  \\
                &&$\vert7\rangle$&$1561.39$&$1607.98$&$(0.61,-0.01,0.21,0.54,-0.51,0.15)$&&$1515.66$&$1588.58$&$(0.55,0.13,-0.43,-0.54,0.43,0.11)$ \\
                &&$\vert8\rangle$&$1245.98$&$1705.71$&$(0.17,-0.11,-0.53,0.58,0.53,-0.23)$&&$1236.29$&$1711.75$&$(-0.20,0.13,-0.56,0.64,0.41,0.23)$\\\hline
				$nn\bar{n}\bar{s}$&	$\frac{1}{2}(2^{+})$&$\vert7\rangle$&$1684.84$&$1587.09$&$(0.34,0.94)$&&$1639.61$&$1592.21$&$(0.61,0.79)$ \\
					&&$\vert8\rangle$&$1600.07$&$1697.62$&$(0.94,-0.34)$& &$1621.03$&$1668.43$&$(0.79,-0.61)$\\
					\hline
				$nn\bar{n}\bar{s}$&	$\frac{3}{2}(0^{+})$&$\vert1\rangle$&$1493.27$&$1027.14$&$(0.64,0.77)$&&$1498.37$&$976.37$&$(-0.61,0.79)$\\
					 &&$\vert7\rangle$&$1344.40$&$1810.53$&$(0.77,-0.64)$& &$1285.66$&$1807.67$&$(0.79,0.61)$\\\hline
				 $nn\bar{n}\bar{s}$&	$\frac{3}{2}(1^{+})$&$\vert3\rangle$&$1328.35$&$1128.27$&$(0.75,-0.63,-0.19)$&&$1294.92$&$1101.95$&$(0.77,0.60,0.20)$\\
					&&$\vert6\rangle$&$1422.54$&$1453.77$&$(0.25,0.01,0.97)$& &$1422.91$&$1403.28$&$(-0.24,0.00,0.97)$\\
					&&$\vert7\rangle$&$1446.70$&$1615.55$&$(0.60,0.78,-0.16)$& &$1395.96$&$1608.56$&$(-0.59,0.79,-0.15)$ \\\hline
				$nn\bar{n}\bar{s}$&$\frac{3}{2}(2^{+})$&$\vert7\rangle$&$1631.45$&&&&$1683.17$&& \\\hline

				$ns\bar{s}\bar{s}$&$\frac{1}{2}(0^{+})$&$\vert1\rangle$&$1949.62$&$1794.42$&$(0.63,0.76)$&&$1965.72$&$1773.32$&$(0.23,0.97)$\\
					 &&$\vert7\rangle$&$1897.46$&$2052.66$&$(0.76,-0.63)$& &$1862.30$&$2054.70$&$(0.97,-0.23)$\\\hline
				$ns\bar{s}\bar{s}$&$\frac{1}{2}(1^{+})$&$\vert4\rangle$&$1790.52$&$1702.83$&$(0.73,0.54,-0.41)$&&$1761.46$&$1680.92$&$(0.76,0.49,-0.42)$\\
					 &&$\vert5\rangle$&$1888.27$&$1890.87$&$(0.68,-0.53,0.50)$& &$1897.84$&$1868.71$&$(0.64,-0.45,0.62)$\\
					 &&$\vert7\rangle$&$1932.54$&$2017.63$&$(-0.05,0.65,0.76)$& &$1894.56$&$2004.23$&$(-0.12,0.74,-0.66)$ \\\hline
				$ns\bar{s}\bar{s}$&$\frac{1}{2}(2^{+})$&$\vert7\rangle$&$1998.37$&&&&$1956.38$&& \\\hline\hline

				\end{longtable}
			\end{sidewaystable}

 \wuhao{\begin{table*}[htbp]
				\caption{A comparison of the calculated masses of $nn\bar{n}\bar{s}$ and $ns\bar{s}\bar{s}$ states for the observed exotic states. Columns 2 and 3 present the results from Model I and II. Columns 4 to 6 list the mass predictions from other representative studies of tetraquark interpretations.}\label{nnns comparison}
				\renewcommand
				\tabcolsep{0.11cm}
				\renewcommand{\arraystretch}{1.0}
				\begin{tabular}{ccccccc}
					\hline\hline
					State&PDG~\cite{ParticleDataGroup:2024cfk}(MeV)&Model I(MeV)&Model II(MeV)&Ref.~\cite{Zhao:2021jss} (MeV)&Ref.~\cite{Deng_2012}(MeV)&Ref.~\cite{Ebert:2008id}(MeV)\\\hline\hline
	$K_0^*(1430)$&1425&1448&1430&$\diagup$&1380&1332\\
    $K_1(1270)$&1253&1174&1146&$\diagup$&1233&1057\\
    $K_1(1400)$&1403&1405&1377&$\diagup$&1456&$\diagup$\\
    $K_1(1650)$&1650&1703&1681&$\diagup$&1644&1855\\
    $K_0^*(1950)$&1957&2053&2055&$\diagup$&1968&$\diagup$\\
    $K_2^*(1980)$&1990&1998&1956&$\diagup$&1968&2001\\
					\hline\hline
				\end{tabular}
			\end{table*}}
		
			For the $nn\bar{n}\bar{s}$ system, four distinct energy levels are predicted in the $I(J^{P})=\frac{1}{2}(0^{+})$ sector. The second-highest energy predictions are 1448.06~MeV and 1429.79 MeV. These values suggest that this state is a plausible candidate for $K^{*}_{0}(1430)$, whose mass and width are $1425\pm50$~MeV and $270\pm80$~MeV~\cite{ParticleDataGroup:2024cfk,Estabrooks:1977xe,Martin:1977ki,BaBar:2021fkz,LHCb:2023evz}. In Ref.~\cite{Ebert:2008id}, Ebert $et\,al.$ predict the $nn\bar{n}\bar{s}$ state with $J^{P}=0^{+}$ at 1332~MeV, which aligns with $K^{*}_{0}(1430)$. Meanwhile, Ref.~\cite{Deng_2012} interprets $K^{*}_{0}(1430)$ as a tetraquark with quantum numbers $N^{2S+1}L_{J}=2^1S_0$. Therefore, $K^{*}_{0}(1430)$ may exhibit a hybrid structure incorporating both $1S$ and $2S$ tetraquark components.

            In the $I(J^{P})=\frac{1}{2}(1^{+})$ sector, there are six distinct energy levels for $nn\bar{n}\bar{s}$ states and three distinct energy levels for $ns\bar{s}\bar{s}$ states predicted in each model. The second-lowest energy predictions are 1173.66~MeV and 1145.81~MeV. These values suggest that this state is a plausible candidate for $K_{1}(1270)$, whose mass and width are $1253\pm7$~MeV and $101\pm12$~MeV~\cite{ParticleDataGroup:2024cfk,Crennell:1967zz,Astier:1969dt,CLEO:2000nrp,LHCb:2017swu,BESIII:2021qfo}. The third-lowest energy predictions are 1404.91~MeV and 1377.45~MeV. These values suggest that this state is a plausible candidate for $K_{1}(1400)$, whose mass and width are $1403\pm7$~MeV and $174\pm13$~MeV~\cite{ParticleDataGroup:2024cfk,Firestone:1972st,CLEO:2000nrp,LHCb:2021uow}. The second-highest and highest energy predictions of $nn\bar{n}\bar{s}$ and the lowest energy predictions of $ns\bar{s}\bar{s}$ are align well with $K_{1}(1650)$, whose mass and width are $1650\pm50$~MeV and $150\pm50$~MeV~\cite{ParticleDataGroup:2024cfk,ACCMOR:1981yww,Bari-Birmingham-CERN-Milan-Paris-Pavia:1982xtk,Frame:1985ka,LHCb:2021uow}.
            Thus $K_{1}(1650)$ may contain both $nn\bar{n}\bar{s}$ and $ns\bar{s}\bar{s}$ tetraquark components. As shown in Table~\ref{nnns comparison}, in Ref.~\cite{Deng_2012}, Deng $et\,al.$ interpret $K_{1}(1270)$ as a $nn\bar{n}\bar{s}$ tetraquark state with quantum numbers $N^{2S+1}L_{J}=1^3S_1$ and $K_{1}(1400)$ as its radial excitation with quantum numbers $N^{2S+1}L_{J}=2^3S_1$. Incorporating predictions from the two models presented herein, $K_{1}(1270)$ is predicted to dominate as a $nn\bar{n}\bar{s}$ tetraquark component, while $K_{1}(1400)$ may exhibit a mixed structure incorporating both $1S-$ and $2S-$ tetraquark components.
			
			For the $ns\bar{s}\bar{s}$ system, both models predict two states in the $I(J^{P})=\frac{1}{2}(0^{+})$ sector and one state in $I(J^{P})=\frac{1}{2}(2^{+})$. The predictions with the highest energy in the $\frac{1}{2}(0^{+})$ sector are 2052.66~MeV and 2054.70~MeV. These values suggest that this state is a plausible candidate for $K_0^*(1950)$, whose mass and width are $1957\pm14$~MeV and $170\pm50$~\cite{ParticleDataGroup:2024cfk,Aston:1987ir,Anisovich:1997qp,BaBar:2021fkz,LHCb:2023evz}. The predictions in the $\frac{1}{2}(2^{+})$ sector are 1998.37~MeV and 1956.38~MeV. These values suggest that this state is a plausible candidate for $K_2^*(1980)$, whose mass and width are $1990^{+60}_{-50}$~MeV and $348^{+50}_{-30}$~MeV~\cite{ParticleDataGroup:2024cfk,Aston:1986jb,BESIII:2019dme,Belle:2020fbd}. In Ref.~\cite{Deng_2012}, Deng $et\,al.$ interpret $K_0^*(1950)$ as a $nn\bar{n}\bar{s}$ tetraquark state with quantum numbers $N^{2S+1}L_{J}=1^5D_0$ and $K_2^*(1980)$ as a $nn\bar{n}\bar{s}$ tetraquark state with quantum numbers $N^{2S+1}L_{J}=2^1D_2$. Incorporating predictions from the two models presented herein, $K_0^*(1950)$ may contain a mixed structure incorporating both $1S$ $ns\bar{s}\bar{s}$ component and $1D$ $nn\bar{n}\bar{s}$ component, while $K_2^*(1980)$ could exhibit a hybrid composition blending $1S$ $ns\bar{s}\bar{s}$ component and $2D$ $nn\bar{n}\bar{s}$ component.
			Furthermore,  several mesons in $I=\frac{3}{2}$ sector are predicted, see Table~\ref{nnns states}.
			
			\section{Summary and Discussion.}\label{General Discussion and Conclusion}

               Investigations on the structure and properties of the light tetraquark states have a long history, starting from 1970s. Theoretically, spectrum of the $qq\bar{q}\bar{q}$ states have been studied using different approaches, such as vairous of quark model, QCD sum rule, lattice QCD.

			In present work, the non-relativistic quark model is adapted to study the spectrum of low-lying light tetraquark states, where two confinement potential schemes and the instanton-induced interaction are employed. The numerical results are consistent with several physical light exotic hadron candidates, indicating that the corresponding compact tetraquark components may contribute to mesons.

            For the $nn\bar{n}\bar{n}$ system, both models anticipate that $f_{0}(500)$ may contain a compact tetraquark component with $I^G(J^{PC})=0^{+}(0^{++})$. One state characterized by the quantum numbers $I^G(J^{PC})=0^{-}(1^{+-})$ is energetically located in close proximity to the resonance $h_{1}(1170)$. Two states with $I^G(J^{PC})=0^{+}(2^{++})$ are predicted to coincide with the resonances $f_{2}(1430)$ and $f_{2}(1640)$, respectively. Furthermore, two states with $I^G(J^{PC})=1^{-}(1^{++})$ are expected to align with $a_{1}(1260)$ and $a_{1}(1640)$, and a state with $I^G(J^{PC})=1^{-}(2^{++})$ may relate to $a_{2}(1700)$, respectively.

            For the $ns\bar{n}\bar{s}$ system, both models anticipate that four states characterized by the quantum numbers $I^G(J^PC)=0^{+}(0^{++})$ are predicted in close proximity to the masses of $f_{0}(1370)$, $f_{0}(1500)$, $f_{0}(2020)$ and $f_{0}(2200)$, respectively. Three states with $I^G(J^{PC})=0^{-}(1^{+-})$ are predicted to coincide with the masses of $h_{1}(1595)$, $h_{1}(1900)$, and $h_{1}(1965)$, respectively. Two states with $I^G(J^{PC})=0^{+}(2^{++})$ are predicted to coincide with the masses $f_{2}(1810)$ and $f_{2}(2010)$, respectively. Similarly, three states possessing $I^G(J^{PC})=1^{-}(0^{++})$ are expected to correlate with $a_{0}(980)$, $a_{0}(1450)$, and $a_{0}(1950)$. Moreover, in the $ss\bar{s}\bar{s}$ system, it is predicted that a single $0^-(1^{+-})$ state aligns with $h_1(2215)$ and a single $0^+(2^{++})$ state responds to $f_2(2150)$.

			The investigation identifies a $nn\bar{n}\bar{s}$ tetraquark state with quantum numbers $I(J^P)=\frac{1}{2}(0^{+})$ aligned with $K^{*}_{0}(1430)$ in the $nn\bar{n}\bar{s}$ system, along with two $nn\bar{n}\bar{s}$ tetraquark states with $I(J^P)=\frac{1}{2}(1^{+})$ corresponding to $K_{1}(1270)$ and $K_{1}(1400)$. Meanwhile, in the $ns\bar{s}\bar{s}$ system, both models predict a single state with $I(J^P)=\frac{1}{2}(0^{+})$ near $K^{*}_{0}(1950)$ and a single state with $I(J^P)=\frac{1}{2}(2^{+})$ responds to $K^{*}_{2}(1980)$. Moreover, $K_{1}(1650)$ may contain both $nn\bar{n}\bar{s}$ and $ns\bar{s}\bar{s}$ tetraquark components with $I(J^{P})=\frac{1}{2}(1^{+})$.

			%It is inferred from these findings that the compact tetraquark states under consideration are likely to possess significant probabilities within the aforementioned exotic hadronic states.
            Additionally, several predictions have not been detected yet. The isotensor state in the window of $1500-1700\,\rm{MeV}$ might be very interesting, and is potentially to be observed in $4\pi$ invariant mass distributions, which deserves further studies both from theoretical and physical sides.

            In summary, the low-lying tetraquark spectrum, with $J^P=0^+,\,1^+,\,2^+$, is explored in the non-relativistic quark model including the Cornell potential (Model I) and instanton interaction in the linear potential (Model II), where the pseudoscalar meson exchange is introduced as the long range interaction between quarks. The analysis of the pattern of the spectrum provides a hint to understand the observations and inspires studies on the light meson spectrum.

            In addtion, one may notice that the instanton-induced interaction can provide us not only the information on mass splitting of the studied teraquark states, but also the coupling between traditional $q\bar{q}$ states and the tetraquark states, this kind of coupling should lead to mixing of $q\bar{q}$ and $qq\bar{q}\bar{q}$ states. Corresponding investigations are now under consideration.

\acknowledgments

This research is supported by the Fundamental Research Funds for the Central Universities under Grant
No. SWU-XDJH202304 and No. SWU-KQ25016, the National Natural Science Foundation of China under Grants No. 12305096.

           \begin{appendix}
           \clearpage
           \section{Expressions of Flavor Wave Functions}\label{Flavor Wave Functions}
           \subsubsection{Flavor Wave Functions of $nn\bar{n}\bar{n}$}
For the $nn\bar{n}\bar{n}$ system ($n$ stands for $u, d$ quarks), there are sixteen flavor configurations for $nn\bar{n}\bar{n}$, that are $uu\bar{u}\bar{u}$, $uu\bar{u}\bar{d}$, $uu\bar{d}\bar{u}$, $ud\bar{u}\bar{u}$, $du\bar{u}\bar{u}$, $uu\bar{d}\bar{d}$, $ud\bar{u}\bar{d}$, $du\bar{u}\bar{d}$, $ud\bar{d}\bar{u}$, $du\bar{d}\bar{u}$, $dd\bar{u}\bar{u}$, $ud\bar{d}\bar{d}$, $du\bar{d}\bar{d}$, $dd\bar{u}\bar{d}$, $dd\bar{d}\bar{u}$, $dd\bar{d}\bar{d}$, which are classified into isospin space, 	
\begin{itemize}
	\item $I=0.$
\begin{align}
	([nn]_{I=0}[\bar{n}\bar{n}]_{I=0})^{0}&=\frac{1}{2}(ud\bar{u}\bar{d}-du\bar{u}\bar{d}-ud\bar{d}\bar{u}+du\bar{d}\bar{u}),\\
    ([nn]_{I=1}[\bar{n}\bar{n}]_{I=1})^{0}&=\frac{1}{\sqrt{12}}(2uu\bar{u}\bar{u}+ud\bar{u}\bar{d}+ud\bar{d}\bar{u}+du\bar{u}\bar{d}+du\bar{d}\bar{u}+2dd\bar{d}\bar{d}).
	\end{align}
	\item $I=1.$
	\begin{align}
		\left([nn]_{I=1}[\bar{n}\bar{n}]_{I=0}\right)^{+}&=\frac{1}{\sqrt{2}}(uu\bar{u}\bar{d}-uu\bar{d}\bar{u}),\\		
		([nn]_{I=1}[\bar{n}\bar{n}]_{I=0})^{0}&=\frac{1}{2}(ud\bar{u}\bar{d}-ud\bar{d}\bar{u}+du\bar{u}\bar{d}-du\bar{d}\bar{u}),\\		
		([nn]_{I=1}[\bar{n}\bar{n}]_{I=0})^{-}&=\frac{1}{\sqrt{2}}(dd\bar{u}\bar{d}-dd\bar{d}\bar{u}),\\		
		([nn]_{I=0}[\bar{n}\bar{n}]_{I=1})^{+}&=\frac{1}{\sqrt{2}}(ud\bar{u}\bar{u}-du\bar{u}\bar{u}),\\
		([nn]_{I=0}[\bar{n}\bar{n}]_{I=1})^{0}&=-\frac{1}{2}(ud\bar{u}\bar{d}+ud\bar{d}\bar{u}-du\bar{u}\bar{d}-du\bar{d}\bar{u}),\\
		([nn]_{I=0}[\bar{n}\bar{n}]_{I=1})^{-}&=\frac{1}{\sqrt{2}}(ud\bar{d}\bar{d}-du\bar{d}\bar{d}),\\
		([nn]_{I=1}[\bar{n}\bar{n}]_{I=1})^{+}&=-\frac{1}{2}(uu\bar{u}\bar{d}+uu\bar{d}\bar{u}+ud\bar{d}\bar{d}+du\bar{d}\bar{d}),\\
		([nn]_{I=1}[\bar{n}\bar{n}]_{I=1})^{0}&=\frac{1}{\sqrt{2}}(uu\bar{u}\bar{u}-dd\bar{d}\bar{d}),\\
		([nn]_{I=1}[\bar{n}\bar{n}]_{I=1})^{-}&=\frac{1}{2}(ud\bar{u}\bar{u}+du\bar{u}\bar{u}+dd\bar{u}\bar{d}+dd\bar{d}\bar{u}).
	\end{align}
	\item $I=2.$
	\begin{align}
		([nn]_{I=1}[\bar{n}\bar{n}]_{I=1})^{++}&=uu\bar{d}\bar{d},\\	([nn]_{I=1}[\bar{n}\bar{n}]_{I=1})^{+}&=-\frac{1}{2}(uu\bar{u}\bar{d}+uu\bar{d}\bar{u}-ud\bar{d}\bar{d}-du\bar{d}\bar{d}),\\
		([nn]_{I=1}[\bar{n}\bar{n}]_{I=1})^{0}&=\frac{1}{\sqrt{6}}(uu\bar{u}\bar{u}-ud\bar{u}\bar{d}-ud\bar{d}\bar{u}-du\bar{d}\bar{u}-du\bar{u}\bar{d}+dd\bar{d}\bar{d}),\\
		([nn]_{I=1}[\bar{n}\bar{n}]_{I=1})^{-}&=\frac{1}{2}(ud\bar{u}\bar{u}+du\bar{u}\bar{u}-dd\bar{u}\bar{d}-dd\bar{d}\bar{u}),\\
		([nn]_{I=1}[\bar{n}\bar{n}]_{I=1})^{--}&=dd\bar{u}\bar{u}.
	\end{align}
\end{itemize}

\subsubsection{Flavor Wave Functions of $ns\bar{n}\bar{s}$ and $ss\bar{s}\bar{s}$   }
			For the $ns\bar{n}\bar{s}$ and $ss\bar{s}\bar{s}$ systems ($n$ stands for $u, d$ quarks), there are four flavor configurations with respect to $ns\bar{n}\bar{s}$. The tetraquark states are classified as,
			\begin{itemize}
				\item $I=0.$
				\begin{align}
					([ns][\bar{n}\bar{s}])^{0}&=\frac{1}{\sqrt{2}}(us\bar{u}\bar{s}+ds\bar{d}\bar{s}).
				\end{align}
				\item $I=1.$
				\begin{align}
					([ns][\bar{n}\bar{s}])^{+}&=-us\bar{d}\bar{s},\\
					([ns][\bar{n}\bar{s}])^{0}&=\frac{1}{\sqrt{2}}(us\bar{u}\bar{s}-ds\bar{d}\bar{s}),\\
					([ns][\bar{n}\bar{s}])^{-}&=ds\bar{u}\bar{s}.
				\end{align}
			\end{itemize}

			\subsubsection{Flavor Wave Functions of $nn\bar{n}\bar{s}$ and $ns\bar{s}\bar{s}$ }
			Note that $[ns][\bar{n}\bar{n}]$, $[ss][\bar{n}\bar{s}]$ are the antiparticles of $nn\bar{n}\bar{s}$, $ns\bar{s}\bar{s}$ respectively. Therefore, we just consider the $nn\bar{n}\bar{s}$ and $ns\bar{s}\bar{s}$ systems in the strangeness $S=1$ sector. There are eight configurations with respect to $nn\bar{n}\bar{s}$, namely $uu\bar{u}\bar{s}$, $uu\bar{d}\bar{s}$, $ud\bar{u}\bar{s}$, $du\bar{u}\bar{s}$, $ud\bar{d}\bar{s}$, $du\bar{d}\bar{s}$, $dd\bar{u}\bar{s}$, $dd\bar{d}\bar{s}$, and two for $ns\bar{s}\bar{s}$, that are $us\bar{s}\bar{s}$,  $ds\bar{s}\bar{s}$. The tetraquark states are grouped into $I=1/2$ and $3/2$ sectors,
			\begin{itemize}
				\item $I=\frac{1}{2}.$
				\begin{align}
					([nn]_{I=0}[\bar{n}\bar{s}]_{I=\frac{1}{2}})^{+}&=-ud\bar{d}\bar{s},\\
					([nn]_{I=0}[\bar{n}\bar{s}]_{I=\frac{1}{2}})^{0}&=ud\bar{u}\bar{s},\\
					([nn]_{I=1}[\bar{n}\bar{s}]_{I=\frac{1}{2}})^{+}&=\sqrt{\frac{2}{3}}uu\bar{u}\bar{s}+\sqrt{\frac{1}{3}}ud\bar{d}\bar{s},\\
			    	([nn]_{I=1}[\bar{n}\bar{s}]_{I=\frac{1}{2}})^{0}&=\sqrt{\frac{1}{3}}ud\bar{u}\bar{s}+\sqrt{\frac{2}{3}}dd\bar{d}\bar{s},\\
			    	([ns]_{I=\frac{1}{2}}[\bar{s}\bar{s}]_{I=0})^{+}&=us\bar{s}\bar{s},\\
			    	([ns]_{I=\frac{1}{2}}[\bar{s}\bar{s}]_{I=0})^{0}&=ds\bar{s}\bar{s}.
				\end{align}
				\item $I=\frac{3}{2}.$
				\begin{align}
					([nn]_{I=1}[\bar{n}\bar{s}]_{I=\frac{1}{2}})^{++}&=-uu\bar{d}\bar{s},\\
					([nn]_{I=1}[\bar{n}\bar{s}]_{I=\frac{1}{2}})^{+}&=\sqrt{\frac{1}{3}}uu\bar{u}\bar{s}-\sqrt{\frac{2}{3}}ud\bar{d}\bar{s},\\
					([nn]_{I=1}[\bar{n}\bar{s}]_{I=\frac{1}{2}})^{0}&=\sqrt{\frac{2}{3}}ud\bar{u}\bar{s}-\sqrt{\frac{1}{3}}dd\bar{d}\bar{s},\\
					([nn]_{I=1}[\bar{n}\bar{s}]_{I=\frac{1}{2}})^{-}&=dd\bar{u}\bar{s}.
				\end{align}
			\end{itemize}
 \clearpage

           \end{appendix}

            %  \bibliography{ref.bib}

\end{document}